\newcommand\beq{\begin{eqnarray}}
\newcommand\eeq{\end{eqnarray}}

\newcommand\missET{E_T^{\rm miss}}
\newcommand\ETmiss{E_T^{\rm miss}}
\newcommand\nS{n_S}
\newcommand\GSM{G_{\rm SM}}
\newcommand\MSbar{\overline{\rm MS}}

\def\lsim{\mathrel{\rlap{\lower4pt\hbox{$\sim$}}
    \raise1pt\hbox{$<$}}}                
\def\gsim{\mathrel{\rlap{\lower4pt\hbox{$\sim$}}
    \raise1pt\hbox{$>$}}}

\documentclass[
amsmath,
prd,nofootinbib,floatfix,12pt,
]{revtex4}

\allowdisplaybreaks
\usepackage{setspace,graphicx}
\usepackage[usenames]{color}

\begin{document}
\renewcommand{\theequation}{\arabic{section}.\arabic{equation}}

\title{\large%
\baselineskip=21pt
Quirks in supersymmetry with gauge coupling unification}

\author{Stephen P. Martin}
\affiliation{
{\it Department of Physics, Northern Illinois University, DeKalb IL 60115,} 
and 
\\
{\it Fermi National Accelerator Laboratory, P.O. Box 500, Batavia IL 60510.}
}

\begin{abstract}\normalsize 
\baselineskip=16pt
\noindent 
I investigate the phenomenology of supersymmetric models with extra 
vector-like supermultiplets that couple to the Standard Model gauge 
fields and transform as the fundamental representation of a new confining 
non-Abelian gauge interaction. If perturbative gauge coupling unification 
is to be maintained, the new group can be $SU(2)$, $SU(3)$, or $SO(3)$. 
The impact on the sparticle mass spectrum is explored, with particular 
attention to the gaugino mass dominated limit in which the supersymmetric 
flavor problem is naturally solved. The new confinement length scale is 
astronomical for $SO(3)$, so the new particles are essentially free. For 
the $SU(2)$ and $SU(3)$ cases, the new vector-like fermions are quirks; 
pair production at colliders yields quirk-antiquirk states bound by 
stable flux tubes that are microscopic but long compared to the new 
confinement scale. I study the reach of the Tevatron and LHC for the 
optimistic case that in a significant fraction of events the 
quirk-antiquirk bound state will lose most of its energy before 
annihilating as quirkonium.
\end{abstract}


\maketitle

\baselineskip=14.9pt
\tableofcontents

\vfill\eject
\baselineskip=14.9pt

\setcounter{footnote}{1}
\setcounter{page}{2}
\setcounter{figure}{0}
\setcounter{table}{0}

\section{Introduction}
\label{sec:intro}
\setcounter{equation}{0}
\setcounter{footnote}{1}

Among the hurdles that must be cleared by any proposed extension of the 
Standard Model (SM) are the stringent limits on quantum 
corrections to the electroweak vector boson propagators due to new physics
\cite{Peskin:1991sw}-\cite{Altarelli:1990zd}. Low-energy supersymmetry \cite{primer}
is generally safe in this regard, because of the fact that all of the new particles 
it introduces get their masses primarily from bare mass terms, 
not from their couplings to the Higgs vacuum expectation values (VEVs).
This includes the Higgs chiral supermultiplets $H_u$ and $H_d$ themselves, 
which are vector-like, 
together forming a self-conjugate representation of the SM
gauge group $\GSM = SU(3)_c \times SU(2)_L \times U(1)_Y$. 
It is therefore interesting to consider 
non-minimal supersymmetric models that maintain this feature by including 
more chiral supermultiplets transforming as vector-like
representations of the gauge group. 

Another well-known and appealing feature of the minimal supersymmetric 
standard model (MSSM) is the perturbative unification of running gauge 
couplings near $M_U \approx 2 \times 10^{16}$ GeV. A sufficient (but not 
necessary) condition for extensions of the MSSM with extra vector-like 
supermultiplets to maintain gauge coupling unification is that the new 
fields come in complete multiplets of the $SU(5)$ global symmetry group 
that contains $\GSM$. This paper studies the properties of models of this 
type that introduce a new non-Abelian gauge group $G_X$, under which the 
new chiral supermultiplets also transform but the MSSM fields are 
neutral. Models of this type have already been introduced by Babu, 
Gogoladze, and Kolda in \cite{Babu:2004xg}, in the context of finding new 
large contributions to the lightest Higgs boson mass.\footnote{Models 
with the same motivation, but without the new non-Abelian gauge group, 
have been studied in \cite{Moroi:1991mg}-\cite{Martin:2010dc}. Other 
recent proposals for extra vector-like chiral supermultiplets are found 
in \cite{othervectorlike}.} I will assume that the new chiral 
supermultiplets have masses at the TeV scale or below, and that the new 
gauge coupling $g_X$ unifies with the $\GSM$ couplings $g_1$, $g_2$, 
$g_3$ at $M_U$. In order to avoid a strong disruption of the running of 
the $\GSM$ gauge couplings, it is necessary that the corresponding 
confinement scale $\Lambda$ for the $G_X$ interactions is below the 
masses of the new fermions and scalars that are also charged under 
$\GSM$. This in turn implies an intriguing phenomenology studied first by 
Okun \cite{Okun:1980kw}, later by Gupta and Quinn \cite{Gupta:1981ve} and 
by Strassler and Zurek \cite{Strassler:2006im}, and more recently in 
considerable depth by Kang and Luty in \cite{Kang:2008ea}. The new 
particles that transform non-trivially under the new gauge group (dubbed 
``theta particle" by Okun, and renamed ``quirks" by Kang and Luty) can 
form exotic bound states with unusual signatures that depend strongly on 
the $G_X$ confinement scale. When a heavy quirk-antiquirk pair is 
produced in a collider experiment, they fly apart but remain connected by 
a stable flux tube, which cannot break due to the large energy cost to 
produce an additional quirk-antiquirk pair. The maximum length of this 
flux tube is roughly of order $L \sim \Delta E/\Lambda^2$, where $\Delta 
E$ is the kinetic energy of the hard scattering production process. This 
length can range from microscopic to literally astronomical, but in any 
case it is much larger than the flux tube thickness $\sim\Lambda^{-1}$. 
The resulting collider signatures are potentially distinctive but also 
possibly quite difficult \cite{Kang:2008ea}-\cite{Abazov:2010yb}.

In this paper, I will study the basic properties of models that maintain 
perturbative unification of gauge couplings, and their renormalization 
group running, in Section \ref{sec:models}. The sparticle mass spectra 
are studied in Section \ref{sec:spectrum}, and Section 
\ref{sec:hierarchy} considers the impact on the supersymmetric little 
hierarchy problem. Some salient aspects of the collider phenomenology of 
the quirks are discussed in Section \ref{sec:pheno}.

\vspace{-0.3cm}

\section{MSSM extended by vector-like 
fields coupled to a new confining non-abelian gauge interaction}
\label{sec:models}
\setcounter{equation}{0}

In order to maintain perturbative gauge coupling unification, the number 
of new particles transforming under the SM gauge group is limited to the 
equivalent of three copies of the ${\bf 5} + {\bf \overline 5}$ of the 
$SU(5)$ group that contains $\GSM$, if they are not much heavier than 1 
TeV. This assumes that $\alpha_i = g_i^2/4\pi$ $(i=1,2,3)$ are required to be 
perturbative (less than 0.3 or so) at and below the energy scale where they 
unify.\footnote{It is crucial to use two-loop (or higher) beta 
functions to correctly implement this perturbativity requirement.
This paper uses three-loop beta functions for supersymmetric gauge couplings
and gaugino masses and two-loop beta functions for Yukawa couplings,
scalar masses, and scalar cubic couplings. These can be found 
straightforwardly from general results in 
refs.~\cite{betas:1,betas:2}, and 
so are not listed explicitly here.}
(One 
could consider unification with larger couplings at and near the 
unification scale, but
then both renormalization group (RG) running and 
threshold corrections will be necessarily out of control, and the 
low-energy manifestation of apparent
unification must be considered merely 
accidental.) It follows that the new gauge non-Abelian 
group must be $G_X = SU(2)_X$ 
or $SO(3)_X$ or $SU(3)_X$. In the following, the new fields are taken to 
transform in the $N=2$, $3$, or $3$ dimensional representations 
respectively for these three cases. Thus the new quirk chiral 
supermultiplets transform under $SU(2)_X \times \GSM$ as:
\vspace{-0.5cm}
\beq
D, \overline D 
&=& ({\bf 2}, {\bf 3}, {\bf 1}, -\frac{1}{3})
+ ({\bf 2}, {\bf \overline 3}, {\bf 1}, \frac{1}{3})
\label{eq:mainDSU2}
\\
\overline L, L
&=& ({\bf 2}, {\bf 1}, {\bf 2}, \frac{1}{2})
+ ({\bf 2}, {\bf 1}, {\bf 2}, -\frac{1}{2})
\label{eq:mainLSU2}
\\
S, \overline S
&=& ({\bf 2}, {\bf 1}, {\bf 1}, 0) \times 2 \nS ,
\label{eq:mainSSU2}
\eeq
\vspace{-0.9cm}

\noindent or under $SU(3)_X \times \GSM$ as:
\vspace{-0.5cm}
\beq
D, \overline D 
&=& ({\bf 3}, {\bf 3}, {\bf 1}, -\frac{1}{3})
+ ({\bf \overline 3}, {\bf \overline 3}, {\bf 1}, \frac{1}{3})
\label{eq:mainDSU3}
\\
\overline L, L
&=& ({\bf 3}, {\bf 1}, {\bf 2}, \frac{1}{2})
+ ({\bf \overline 3}, {\bf 1}, {\bf 2}, -\frac{1}{2})
\label{eq:mainLSU3}
\\
S, \overline S
&=& [({\bf 3}, {\bf 1}, {\bf 1}, 0)
+ ({\bf \overline 3}, {\bf 1}, {\bf 1}, 0)] \times \nS ,
\label{eq:mainSSU3}
\eeq
\vspace{-0.9cm}

\noindent or under $SO(3)_X \times \GSM$ as:
\vspace{-0.5cm}
\beq
D, \overline D 
&=& ({\bf 3}, {\bf 3}, {\bf 1}, -\frac{1}{3})
+ ({\bf 3}, {\bf \overline 3}, {\bf 1}, \frac{1}{3})
\label{eq:mainDSO3}
\\
\overline L, L
&=& ({\bf 3}, {\bf 1}, {\bf 2}, \frac{1}{2})
+ ({\bf 3}, {\bf 1}, {\bf 2}, -\frac{1}{2})
\label{eq:mainLSO3}
\\
S \equiv \overline S &=& ({\bf 3}, {\bf 1}, {\bf 1}, 0) \times \nS .
\label{eq:mainSSO3}
\eeq
\vspace{-0.8cm}

\noindent These are the main model frameworks considered below. 
With these assignments, $D, \overline L$ transform as a ${\bf 5}$ and 
$\overline D, L$ transform as a ${\bf \overline 5}$ of the usual 
Georgi-Glashow $SU(5)$, ensuring that the unification of $\GSM$ gauge couplings 
persists. Also included are $\nS$ $\GSM$ 
singlets in the same representations 
of $G_X$. Note that since $SU(2)_X$ and $SO(3)_X$ have the same Lie 
algebra, the practical distinction between them is really whether the 
representations of the chiral superfields are doublets or triplets.

[There are some variations on the above models that are consistent with 
gauge coupling unification with the new fields at the TeV scale, which 
should be mentioned although they are inconsistent with an assignment of
$D, \overline D, L, \overline L$ into ${\bf 5} + {\bf \overline 5}$ of $SU(5)$. First, for $SU(3)_X$ only, there is another, 
inequivalent, embedding in which $D, \overline D$ have the same 
assignments, but $\overline L, L = ({\bf \overline 3}, {\bf 1}, {\bf 2}, 
\frac{1}{2}) + ({\bf 3}, {\bf 1}, {\bf 2}, -\frac{1}{2}) $ instead. 
Also, for $SU(3)_X$, one could put $D, \overline D$ into 
three singlets if $L, \overline L$ are in a triplet, or vice versa. 
Likewise, for $SU(2)_X$ or $SO(3)_X$, one could put $D, \overline D$ into any 
combination 
of $n^D_1$ singlets, $n^D_2$ doublets, and $n^D_3$ triplets, 
and similarly for $\overline L, L$, provided that $n^D_1 + 2 n^D_2 + 3 
n^D_3 = n^L_1 + 2 n^L_2 + 3 n^L_3 \leq 3$. Finally, it should be noted 
that the number and type of $G_X$ representations of the SM singlets do not 
affect gauge coupling unification for $\GSM$, and so are more 
generally arbitrary as 
long as they are anomaly-free under $G_X$. 
However, to keep the discussion below 
bounded, I will limit the discussion below to the models defined 
by eqs.~(\ref{eq:mainDSU2})-(\ref{eq:mainSSO3}).]

The supersymmetric mass parameters of the $D, \overline D, L, \overline L$ 
fields are assumed to arise by the same mechanism that gives the entirely 
analogous term $\mu H_u H_d$ in the superpotential of the MSSM. 
For example \cite{Kim:1983dt,Murayama:1992dj}, 
one may assume that the mass terms $H_u H_d$ and $D 
\overline D$ and $L \overline L$ and $S \overline S$ are forbidden at 
tree-level, and arise from non-renormalizable superpotential terms:
\beq
W = \frac{1}{M_P^2} X \overline X  \left (
\lambda_\mu H_u H_d + \lambda_D D \overline D + \lambda_L L \overline L + 
\lambda_{S_i} S_i 
\overline S_i \right )
\label{eq:axionicmasses}
\eeq
(with an implied sum over $i=1,\ldots, \nS$ if $\nS \not= 0$)
when the fields $X, \overline X$ get VEVs roughly of order 
$10^{11}$ GeV. Here 
$M_P = 2.4 \times 10^{18}$ GeV is the reduced Planck mass. These 
intermediate-scale VEVs are natural,
for example \cite{Murayama:1992dj}, if there is also a superpotential
\beq
W = \frac{\lambda_X}{4 M_P^2} X^3 \overline X
\eeq
and soft terms
\beq
-{\cal L}_{\rm soft} = m_X^2 |X|^2 + m_{\overline X}^2 |\overline X|^2
+ \left (\frac{a_X}{4 M_P^2} X^3 \overline X + {\rm c.c.} \right ).
\eeq
Non-trivial VEVs for $X, \overline X$ break a Peccei-Quinn symmetry,
giving rise to an invisible axion solution to the strong CP problem
\cite{Kim:1983dt}.
There will be a non-trivial 
local minimum of the potential provided that 
$|a_X|^2 - 6 |\lambda_X|^2 (m_X^2 + m_{\overline X}^2) > 0$, 
and it will be a global minimum if 
$|a_X|^2 - 8 |\lambda_X|^2 (m_X^2 + m_{\overline X}^2) > 0$
\cite{minX}.
This will give rise to the vector-like mass terms in the low-energy 
effective superpotential
\beq
W = \mu H_u H_d 
+ \mu_D D \overline D + \mu_L L \overline L + \mu_{S_i} S_i \overline S_i
.
\label{eq:masses}
\eeq
with $\mu, \mu_D, \mu_L, \mu_S$ of order 100 GeV to 1 TeV, provided that 
the corresponding couplings $\lambda_\mu, \lambda_D, \lambda_L, 
\lambda_S$ are not too small.

The $S, \overline S$ fields do not couple to $\GSM$ gauge fields, and so 
are not constrained by LEP2 or Tevatron or other direct production, nor 
do they affect the SM gauge couplings directly. So, some number $n$ of 
them (with $0 \leq n \leq n_S$) could actually have current masses 
$\mu_{S_i}$ that are far below the electroweak scale. This would occur if 
the $\lambda_{S_i}$ coupling(s) in eq.~(\ref{eq:axionicmasses}) are 
absent (perhaps replaced by terms of even higher dimensionality), or just 
small. Note that for $n>0$, there will be no stable flux tubes for 
pair-produced particles charged under $G_X$, because then as the 
particles produced in the hard collision fly apart, the gauge string will 
break to form bound states with size of order $\Lambda^{-1}$ just as in 
ordinary QCD. This is because the energy cost to produce an additional 
pair of light $S, \overline S$ after the hard collision would then be 
small.

The new fermion content of the theory consists of 
a color triplet charge $\pm 1/3$ Dirac fermion 
$(\psi_D, {\overline \psi_D})$ with mass $\mu_D$;
a charge $\pm 1$ 
Dirac fermion
$(\psi^+_{\overline L}, \psi^-_{L})$ with mass $\mu_L$;
and charge 0
fermions $\psi_{\overline L}^0, \psi_L^0, \psi_{S_i}^0, 
\psi_{\overline S_i}^0$. 
The scalar partners of these particles
will have soft-supersymmetry breaking squared-mass terms:
\beq
-{\cal L} &=& 
m_{\widetilde D}^2 |D|^2 + 
m_{\widetilde{\overline{D}}}^2 |\overline{D}|^2 + 
m_{\widetilde L}^2 |L|^2 + 
m_{\widetilde{\overline{L}}}^2 |{\overline L}|^2 + 
(m_{\widetilde S}^2)_{ij} S_i^* S_j + 
(m_{\widetilde{\overline{S}}}^2)_{ij} \overline S_i^* \overline S_j + 
\nonumber \\ &&
+ 
( b_D D \overline D + b_L L \overline L + 
(b_{S})_{ij} S_i \overline S_j + {\rm c.c.} ), 
\eeq
where the scalar components are denoted by the same symbol as the 
chiral supermultiplets of which they are members.
In the case $G_X = SU(2)$ or $SO(3)$, the fields $S_i$ and $\overline 
S_i$ actually have the same quantum numbers, and so can mix with further
soft mass terms $(m_{\widetilde S}^2)_{ij} S_i^* \overline S_j$, etc.,
but for simplicity I assume that mixing 
between $S_i$ and $\overline S_j$ chiral supermultiplets is absent.
Also for simplicity, I will assume that the above $S_i$ and $\overline S_j$ soft 
terms are diagonal in the same basis that the superpotential masses 
$\mu_{S_i}$ are diagonal. This is natural if the soft supersymmetry 
breaking arises in a flavor-blind framework such as gaugino mass 
dominance.

In the absence of Yukawa couplings involving the new chiral supermultiplets, 
the charge 0 fermions are unmixed, and form Dirac fermions with 
masses $\mu_L$ and $\mu_{S_i}$. 
For $\nS > 0$, the new chiral supermultiplets can have Yukawa couplings
in addition to their mass terms in eq.~(\ref{eq:masses}):
\beq
W = 
k_i H_u L \overline S_i + k'_i H_d \overline L S_i.
\label{eq:defineYukawas}
\eeq
and corresponding soft scalar cubic
terms,
\beq
-{\cal L} = a_{k_i} H_u L {\overline S_i} 
+ a_{k_i'} H_d {\overline L} {S_i} + {\rm c.c.}
\eeq
As mentioned above, $H_u L S$ and $H_d \overline L 
\overline S$ couplings for $SU(2)_X$ or $SO(3)_X$ are also possible, but 
are omitted here for 
simplicity. The superpotential Yukawa couplings produce mixing between 
the gauge eigenstate fermions, yielding Dirac fermions 
$(\psi_j^0, \overline \psi_j^0)$ which are mixtures of $\psi_L^0, 
\psi_{S_i}^0$ and of $\psi_{\overline L}^0, \psi_{\overline S_i}^0$, 
respectively. For example, if 
only one pair $S$ and $\overline S$ has couplings to the Higgs fields, 
then the new neutral fermion mass matrix becomes
\beq
-{\cal L} &=& \Bigl ( \psi_{L^0} \>\>\, \psi_{S} \Bigr )
\begin{pmatrix}
\mu_L & k v_u \cr
k' v_d & \mu_S
\end{pmatrix}
\begin{pmatrix}
\psi_{\overline L^0} \cr \psi_{\overline S}
\end{pmatrix}
+ {\rm c.c.}
\eeq
when the MSSM 
Higgs fields get their VEVs $v_u, v_d$ with $\tan\beta = v_u/v_d$ and
$v = \sqrt{v_u^2 + v_d^2} \approx 175$ GeV. 
The couplings $k, k'$ gives rise to 1-loop effects that can 
significantly raise 
the lightest Higgs 
scalar boson mass due to a lack of complete cancellation 
between scalar and fermion loops, especially for large 
$k$ if $\tan\beta$ is not small. 
This was the motivation of \cite{Babu:2004xg}, but
as noted in similar contexts in \cite{Babu:2008ge,Martin:2009bg}
and remarked on further below, it is doubtful whether this 
really ameliorates the supersymmetric little hierarchy problem. 

In keeping with the idea that the apparent gauge coupling unification for 
$\GSM$ is telling us something important about the underlying theory, I 
will assume that the new non-Abelian gauge coupling $g_X$ unifies with 
$g_1$, $g_2$ and $g_3$ at a scale $M_U \gsim 2 \times 10^{16}$ GeV. In 
practice, I use three-loop RG equations to run up from the electroweak 
scale, and declare the scale where $g_1 = g'\sqrt{5/3}$ and $g_2$ meet to be $M_U$, and 
require $g_X$ to be equal to them there. The QCD coupling $g_3$ typically 
misses this common value at $M_U$ by a small amount that can be 
reasonably ascribed to threshold corrections. Now, RG running $g_X$ 
from this scale, I require that it remains finite down to scales well below 
the masses of the quirks $D, \overline D, \overline L, L$. Otherwise, 
two-loop effects would strongly affect the running of the SM gauge 
couplings, rendering their apparent unification merely accidental. For 
$SU(2)_X$ and $SO(3)_X$, this requirement is automatically satisfied for 
all $\nS \geq 0$, but for $SU(3)_X$ it requires $\nS \geq 3$. I therefore 
consider $\nS = 3$ to be the minimal viable model for the $SU(3)_X$ case.

For illustration, the running of the gauge couplings is shown at 
three-loop order in Figure 1, for the three cases 
$SU(2)_X$ with $\nS = 0$ and $SO(3)_X$ with 
$\nS = 0$ and $SU(3)_X$ with
$\nS = 3$. For simplicity, I have assumed vanishing Yukawa couplings
and chosen a single scale 
$M_{\rm thresh} = 1$ TeV as the effective average mass of the new 
particles charged under $G_X$ and the MSSM superpartners. The 
unification will have some dependence on
the actual thresholds, which one might imagine is roughly comparable to the 
unknown threshold dependence due to high-scale particles.
\begin{figure}[!tp]
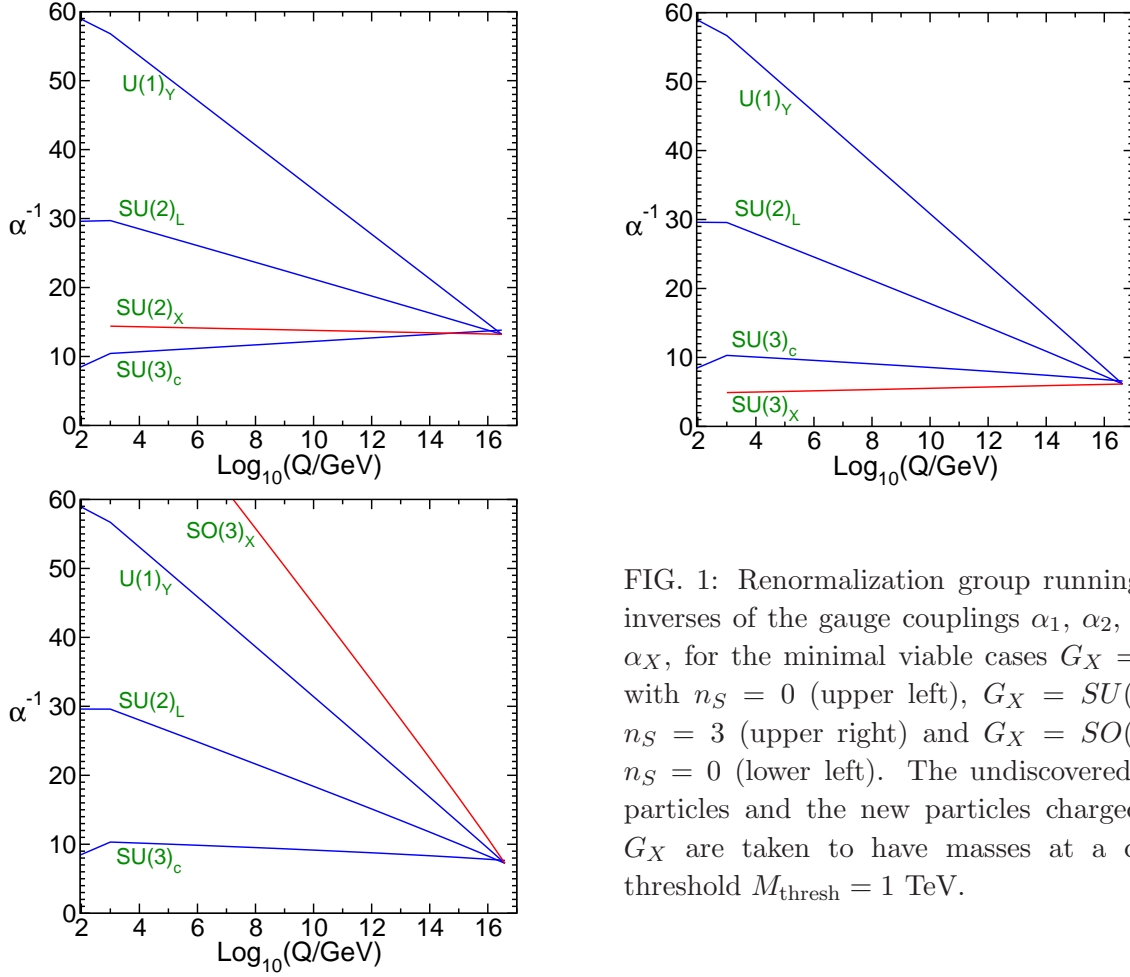

\begin{minipage}[]{0.49\linewidth}
\begin{flushleft}
\includegraphics[width=6.8cm,angle=0]{unifSU2_0.eps}
\\
\includegraphics[width=6.8cm,angle=0]{unifSO3_0.eps}
\end{flushleft}
\end{minipage}
\begin{minipage}[]{0.49\linewidth}
\begin{flushleft}
\includegraphics[width=6.8cm,angle=0]{unifSU3_3.eps}
\hspace{1.5cm}
\caption{\label{fig:alphainvrun}
Renormalization group running of the inverses of the gauge couplings 
$\alpha_1$, $\alpha_2$, $\alpha_3$, and $\alpha_X$, for the minimal 
viable cases $G_X = SU(2)$ with $\nS =0$ (upper left), $G_X = SU(3)$ 
with $\nS =3$ (upper right) and $G_X = SO(3)$ with $\nS =0$ (lower 
left). The undiscovered MSSM particles and the new particles charged 
under $G_X$ are taken to have masses at a common threshold $M_{\rm 
thresh} = 1$ TeV.}
\vspace{0.47cm}\phantom{xxx}
\end{flushleft}
\end{minipage}
\end{figure}
In the case of $SU(2)_X$ with $\nS = 0$, the gauge coupling $g_X$ 
runs quite slowly, 
and is somewhat weaker than the QCD coupling at the TeV scale. In 
contrast, for the $SO(3)_X$ case with $\nS = 0$, $g_X$ runs quickly to very small 
values in the infrared, due to a large positive beta function coefficient. 
For the 
minimal viable $SU(3)_X$ case with $\nS = 3$, the $g_X$ beta function is 
even more negative than the QCD beta function, leading to a gauge 
coupling at the TeV scale that is larger, but still perturbative and not 
running very fast. For non-minimal models with $\nS$ larger than these 
values, the TeV-scale values of $\alpha_X$ are smaller, because the $g_X$ 
beta function is larger.

Below the masses of the quirks and their supersymmetric partners, the
coupling $g_X$ has a negative beta function, and diverges at some scale 
$\Lambda$ when calculated at any particular loop order in a specified scheme. 
Given the
$\MSbar$ beta function for $\alpha_X = g_X^2/4\pi$ up to 4-loop order:
\beq
\beta_{\alpha_X} = Q \frac{d\alpha_X}{dQ} = -2\left (b_0 
\alpha_X^2 +b_1 
\alpha_X^3 + b_2 \alpha_X^4 + b_3 \alpha_X^5 + \ldots \right ),
\eeq
the scale $\Lambda$ can be defined,\footnote{Note that the definition 
for $\Lambda$  used here corresponds to 
$\Lambda/4$ in ref.~\cite{Kang:2008ea}.}
using any convenient $\alpha_X(Q_0)$ with $Q_0 \leq M_{\rm thresh}$
as input,
by an expansion in inverse powers of
$t \equiv \ln(Q_0^2/\Lambda^2)$ \cite{Chetyrkin:1997un}: 
\beq
\alpha_X(Q_0) &=& \frac{1}{b_0 t} \Bigl (1 - [b_1 \ln t]/b_0^2 t
+ [b_0 b_2 + b_1^2 (\ln^2 t - \ln t -1) ]/b_0^4 t^2 
\nonumber \\ &&
+ [b_0^2 b_3 - b_1^3 (2 \ln^3 t - 5 \ln^2 t - 4 \ln t + 1) - 6 b_0 b_1 
b_2 \ln t]/2 b_0^6 t^3  + \ldots\Bigr ).
\label{eq:estimateLambda}
\eeq
It is common in rough estimates to only use the one-loop-order estimate
$\alpha_X(Q_0) = 1/b_0 t$, with $b_0 = (11 C_A  - 2 T_F)/12 \pi$ 
where $(C_A, T_F) = (N, n)$ for
$SU(N)_X$ and $(C_A, T_F) = (2, 2n)$ for $SO(3)_X$, with $n$ denoting the 
number of
the $\nS$ SM singlet fields that have masses below $Q_0$ 
and are treated as non-decoupled below $M_{\rm thresh}$. 
However, it turns out
that including the higher loop effects (with coefficients $b_{1,2,3}$ 
found in refs.~\cite{fourloop}) are quite important for obtaining 
a stable value of the $G_X$ confinement scale 
$\Lambda$. This is illustrated in Table \ref{tab:Lambda},
which shows the results obtained for $\Lambda^{(\ell)}_{\MSbar}$
at various loop orders $\ell$, assuming again that the effective
decoupling scale for particles charged under $G_X$ is 
$M_{\rm thresh} = 1$ TeV.
\renewcommand{\arraystretch}{1.25}
\begin{table}
\begin{tabular}[c]{ccccccc}
$G_X$~~ & ~~$\nS$~ & ~~$\alpha_X^{-1}$(1 TeV)~~ 
& $\Lambda^{(1)}$ & $\Lambda^{(2)}$ & $\Lambda^{(3)}$ & $\Lambda^{(4)}$ 
\\
\hline\hline
$SU(2)$ & 0 & 9.3 & 
~0.35 GeV~ & ~1.3 GeV~ & ~1.1 GeV~ & ~1.1 GeV
\\
\phantom{x} & 1 & 14.4 & 
~4.4 MeV~ & ~19 MeV~ & ~17 MeV~ & ~17 MeV
\\
\phantom{x} & 2 & 19.5 & 
~57 keV~ & ~280 keV~ & ~250 keV~ & ~250 keV
\\
\phantom{x} & 3 & 24.5 & 
~0.76 keV~ & ~4.1 keV~ & ~3.7 keV~ & ~3.7 keV
\\
\phantom{x} & 4 & 29.5 & 
~11 eV~ & ~60 eV~ & ~55 eV~ & ~55 eV
\\
\hline
$SU(3)$ & 3 & 4.9 & 
~61 GeV~ & ~140 GeV~ & ~120 GeV~ & ~120 GeV
\\
 & 4 & 9.9 & 
~3.5 GeV~ & ~11 GeV~ & ~9.3 GeV~ & ~9.5 GeV
\\ 
 & 5 & 15.0 & 
~190 MeV~ & ~720 MeV~ & ~620 MeV~ & ~620 MeV
 \\
 & 6 & 20.1 & 
~10 MeV~ & ~44 MeV~ & ~38 MeV~ & ~38 MeV
 \\
 & 7 & 25.1 & 
~0.59 MeV~ & ~2.7 MeV~ & ~2.4 MeV~ & ~2.4 MeV
 \\
 & 8 & 30.2 & 
~32 keV~ & ~160 keV~ & ~140 keV~ & ~140 keV
 \\
 & 9 & 35.2 & 
~1.9 keV~ & ~9.7 keV~ & ~8.7 keV~ & ~8.8 keV
 \\
\hline
$SO(3)$ & 0 & 83 & 
~$1.3\!\times\! 10^{-19}$ eV~ & ~$1.1\!\times\! 10^{-18}$ eV~ & 
~$1.0\!\times\! 10^{-18}$ eV~ & ~$1.0\!\times\! 10^{-18}$ eV
\\
& 1 & 103 & 
~$4.7\!\times\! 10^{-27}$ eV~ & ~$4.4\!\times\! 10^{-26}$ eV~ & 
~$4.2\!\times\! 10^{-26}$ eV~ & ~$4.2\!\times\! 10^{-26}$ eV
\\
\hline
\end{tabular}
\caption{The $G_X$ confinement scale $\Lambda^{(\ell)}_{\MSbar}$, computed
at various loop orders $\ell$ by using eq.~(\ref{eq:estimateLambda}) 
keeping terms of order $1/t^\ell$. The 
new particles charged under $G_X$ are taken to have an effective average
decoupling mass scale of $Q_0 = M_{\rm thresh} = 1$ TeV, with no SM 
singlet quirks with current masses $\mu_{S_i}$ smaller than $\Lambda$.
\label{tab:Lambda}}
\end{table}
The point of carrying 
the calculation to 4-loop order is {\em not} because of the 
very slightly increased 
accuracy obtained 
(since there are threshold uncertainties here that are not known), 
but rather to demonstrate the stability of the results with respect to 
inclusion of higher-order terms. In fact, the 4-loop order results for $\Lambda$ hardly 
differ at all from the 3-loop order ones, and only at the 10\% level from 
the 2-loop order ones. However, they are notably larger than  the 
1-loop order estimate, which is therefore judged to be deprecated as an
estimate of the physical $G_X$ confinement scale.

Table \ref{tab:Lambda} shows that the confinement scale $\Lambda$ for 
$SO(3)_X$ is very small in energy units. In terms of length, the 
confinement scale for the minimal model $\nS=0$ is of the order $10^{11}$ 
meters, very roughly of order the radius of the Earth's orbit around the 
Sun. For $\nS=1$, the confinement length is of order 100 parsecs. Thus 
for all practical purposes, the quirks are actually free. Adding other SM 
singlets charged under $SO(3)_X$ will only decrease $\Lambda$, making the 
confinement length even larger.

For the minimal viable $SU(2)_X$ and $SU(3)_X$ models, the confinement 
energy scale is much larger. 
Increasing $\nS$ leads to smaller $\Lambda$, as indicated in Table 
\ref{tab:Lambda}.
If $n$ of the $n_S$ SM singlets charged under $G_X$ have current 
masses $\mu_{S_i}$ less than $\Lambda$, 
the confinement scale will be decreased. This is illustrated in 
Table \ref{tab:Lambdalow}
for the extreme case that all $n=n_S$ of the new singlets are lighter 
than $\Lambda$.
\renewcommand{\arraystretch}{1.25}
\begin{table}
\begin{tabular}[c]{cccc}
$G_X$~~ & ~~$\nS = n$~ & ~~$\alpha_X^{-1}$(1 TeV)~~ 
& $\Lambda^{(4)}$ 
\\
\hline\hline
$SU(2)$ & 1 & 14.4 & 
~5.0 MeV
\\
& 2 & 19.5 &
~6.3 keV
\\
& 3 & 24.5 &
~1.3 eV
\\
& 4 & 29.5 &
~$2.0 \times 10^{-5}$ eV
\\
\hline
$SU(3)$ & 3 & 4.9 & 
~68 GeV
\\
 & 4 & 9.9 & 
~1.5 GeV
\\ 
 & 5 & 15.0 & 
~13 MeV
\\
 & 6 & 20.1 & 
~38 keV
\\
 & 7 & 25.1 & ~30 eV
\\
 & 8 & 30.2 & ~0.0030 eV
\\
 & 9 & 35.2 & ~$2.1 \times 10^{-8}$ eV
\\
\hline
\end{tabular}
\caption{As in Table \ref{tab:Lambda}, but now taking all $\nS = n$ of the 
SM singlet
fermions to have current masses much less than $\Lambda$, and showing only
the 4-loop order result (which is nearly identical to the 3-loop order 
result in all cases).
\label{tab:Lambdalow}}
\end{table}
Both Tables \ref{tab:Lambda} and \ref{tab:Lambdalow} take the 
effective average decoupling scale for the particles 
in the other new chiral supermultiplets 
(including both scalars and fermions) to be 1 TeV.
More generally one can estimate:
\beq
\Lambda =
\Lambda_{\rm tab}
\left (\frac{M_{\rm thresh}}{\mbox{TeV}}\right )^{1 - \Delta},\qquad
\Delta = \frac{3 C_A - 5 - \nS}{(11 C_A - 2 n)/3}, 
\label{eq:Lambdarat}
\eeq
where now $\Lambda_{\rm tab}$ is
the value given in Table \ref{tab:Lambda} for $n=0$,
or Table \ref{tab:Lambdalow} for $n=\nS$. Here,
$M_{\rm thresh}$ is defined to be the effective average 
decoupling scale for the new supermultiplets, at which
the threshold corrections to the gauge coupling $g_X$ are small.
As seen in Figure 
\ref{fig:alphainvrun}, 
in the minimal viable models
the $g_X$ coupling runs fairly slowly 
in the non-decoupled theory above $Q = M_{\rm thresh}$. 
This means that in the minimal model for $SU(2)_X$, $\Lambda \approx 
0.001 M_{\rm thresh}$, while for the minimal viable $SU(3)_X$ model with 
$\nS =3$, $\Lambda \approx 0.12 M_{\rm thresh}$, if $n=0$. 

If the Yukawa couplings $k,k'$ are present, there is a potentially important constraint from precision electroweak observables.
The new contributions to the Peskin-Takeuchi $S,T$ observables from the new fermions are:
\beq
\Delta T &=& \frac{N v^4}{480 \pi s_W^2 M_W^2 M_F^2}
[13 (\hat k^4 + \hat k^{\prime 4}) 
+ 2 (\hat k^3 \hat k^{\prime} + \hat k \hat k^{\prime 3})
+ 18 \hat k^2 \hat k^{\prime 2} ],
\\
\Delta S &=& \frac{N v^2}{30 \pi M_F^2}
[4 \hat k^2 + 4 \hat k^{\prime 2} -7 \hat k \hat k^{\prime} ].
\eeq
where $\hat k = k \sin\beta$ and $\hat k' = k' \cos\beta$ and $v \approx 
175$ GeV, and for illustration purposes I have chosen 
$\mu_L \approx \mu_S \gg m_Z$ and 
assumed that the corresponding scalars are much heavier.
The values of these Yukawa couplings are governed by infrared quasi-fixed points. For example, if $k'$ is negligible, then 
the beta functions for $k$ and the top Yukawa 
coupling are given at one-loop order by:
\beq
Q\frac{dy_t}{dQ} &=& \beta_{y_t} = \frac{y_t}{16 \pi^2}
\left [ 6 y_t^2 + y_b^2 + N k^2 - \frac{16}{3} g_3^2 - 3 g_2^2 - 
\frac{13}{15} g_1^2 \right ]
,
\\
Q\frac{dk}{dQ} &=& \beta_{k} = \frac{k}{16 \pi^2}
\left [ (3 + N) k^2 + 3 y_t^2 - 4 C g_X^2 - 3 g_2^2 - 
\frac{3}{5} g_1^2 \right ] ,
\eeq
where $N = 2$, $3$, or $3$ and $C = 3/4$, $4/3$, or $2$ for $G_X = 
SU(2)$, $SU(3)$, or $SO(3)$
respectively. 
The fixed points arise due to the balancing between the positive Yukawa and the negative gauge contributions \cite{PRH}.
Including two-loop effects, I find for the minimal 
viable models the infrared quasi-fixed-point values:
\vspace{-0.4cm}
\beq
k_{\rm fixed} = 
\left \{ \begin{array}{l}
0.88\qquad [SU(2)_X, \nS = 0]
\\[-9pt]
0.76\qquad [SO(3)_X, \nS = 0]
\\[-9pt]
1.32\qquad [SU(3)_X, \nS = 3]
.
\end{array}
\right.
\label{eq:kfixed}
\eeq
\vspace{-0.45cm}

\noindent
at $Q = 1$ TeV. The resulting contributions to $S,T$ can be used to put a lower bound on $\mu_L$. Requiring the results to be within the current 95\% CL ellipse
from experimental results on $m_t$, $m_W$, and $Z$-peak observables using the same methodology as in \cite{Martin:2009bg}, I
estimate $\mu_L > 210, 225, 380$ GeV
for the
$G_X = SU(2), SO(3), SU(3)$ fixed point cases respectively. 
However, $G_X$ confinement may play a significant role 
in modifying this estimate for $SU(3)$,
because $\Lambda$ in that case is larger than $M_Z$. For smaller 
Yukawa couplings $k \ll k_{\rm fixed}$, there is no constraint
as the vector-like particles decouple from precision electroweak observables.

\section{Soft SUSY-breaking masses and the sparticle spectrum}
\label{sec:spectrum}
\setcounter{equation}{0}

The presence of new vector-like supermultiplets has a profound effect on 
the spectrum of superpartner masses. They cause the $\GSM$ gauge 
couplings to run to much larger values in the ultraviolet as they approach 
unification, resulting in bigger one-loop contributions to soft scalar 
squared masses from RG running, compared to the MSSM. The new 
supermultiplets also allow the $G_X$ gaugino masses to contribute 
indirectly to MSSM gaugino and sfermion masses, through two-loop order 
effects. In this section, the patterns of soft supersymmetry breaking 
masses will be considered for these models. For simplicity, the 
discussion will be mostly limited to the scenario in which a unified 
gaugino mass parameter $m_{1/2}$ is much larger than the
scalar masses and other sources of supersymmetry breaking at the RG scale 
where the gauge couplings unify. This gaugino mass dominated limit 
is motivated as a solution to the supersymmetric flavor problem, 
since it automatically produces flavor-blind soft terms.

The modified running of the gaugino masses pushes them to be smaller near 
the TeV scale than they would be in the MSSM. Given an input 
unified gaugino mass $m_{1/2}$ at the unification scale, one finds for 
the running gaugino masses at $Q = 1$ TeV:
\vspace{-0.5cm}
\beq
(M_1,\> M_2,\> M_3) &=& m_{1/2} \times 
\left \{ \begin{array}{ll}
(0.41,\>0.77,\>2.28) 
\quad &\mbox{[MSSM],}
\\[-7pt]
(0.21,\>0.39,\>1.16)
\quad &\mbox{[$SU(2)$, $\nS=0$],}
\\[-7pt]
(0.112,\>0.20,\>0.57) 
\quad &\mbox{[$SO(3)$, $\nS=0$],}
\\[-7pt]
(0.080,\>0.135,\>0.40) 
\quad &\mbox{[$SU(3)$, $\nS=3$].}
\end{array}
\right .
\label{eq:comparegauginos}
\eeq
\vspace{-0.6cm}

\noindent Thus, in the extended models, to obtain the same physical 
gaugino masses, one must start with larger 
$m_{1/2}$ than one would in the MSSM. 
Since $m_{1/2}$ is not directly observable, it is also interesting to 
consider the ratios of these gaugino
masses. They are also affected, but more mildly 
(being due to 2-loop effects):
\vspace{-0.5cm}
\beq
(M_2/M_1,\> M_3/M_2,\> M_3/M_1) &=& 
\left \{ \begin{array}{ll}
(1.87,\>2.96,\>5.53) 
\quad &\mbox{[MSSM],}
\\[-7pt]
(1.85,\>2.95,\>5.44)
\quad &\mbox{[$SU(2)$, $\nS=0$],}
\\[-7pt]
(1.79,\>2.84,\>5.09) 
\quad &\mbox{[$SO(3)$, $\nS=0$],}
\\[-7pt]
(1.73,\>2.85,\>4.95) 
\quad &\mbox{[$SU(3)$, $\nS=3$],}
\end{array}
\right .
\eeq
\vspace{-0.6cm}

\noindent where again unification of gaugino masses at the gauge coupling 
unification scale is assumed. The effect of the additional fields is thus 
to somewhat compress the gaugino mass spectrum compared to the MSSM case, 
with the ratio of gluino to bino masses decreased by about 10 per cent 
for $G_X = SO(3)$ and $SU(3)$. To obtain the physical masses, 
one must also include mixing with Higgsinos and the pole mass corrections, which are 
particularly important for the gluino \cite{MV,PBMZ,MV2}.

In the extended models the 
squark and slepton masses are also relatively smaller 
(compared to $m_{1/2}$) at the TeV scale than in the MSSM. 
Taking a gaugino-mass dominated scenario (by assuming a vanishing common scalar 
squared mass $m_0^2 = 0$ at the unification scale), one finds for the 
first and second family squark and slepton masses at $Q = 1$ TeV:
\vspace{-0.5cm}
\beq
(
m_{\widetilde q_1},\,
m_{\widetilde{\bar u}_1},\,
m_{\widetilde{\bar d}_1},\,
m_{\tilde \ell_1},\,
m_{\widetilde{\bar e}_1})
&=& m_{1/2}\! \times\! 
\left \{ \begin{array}{ll}
(2.15,\, 2.08,\, 2.07,\, 0.67,\, 0.37) 
\quad &\mbox{[MSSM],}
\\[-7pt]
(1.61,\,1.55,\,1.54,\,0.57,\,0.33)
\quad &\mbox{[$SU(2)$, $\nS=0$],}
\\[-7pt]
(1.23,\,1.18,\,1.17,\,0.48,\,0.28)
\quad &\mbox{[$SO(3)$, $\nS=0$],}
\\[-7pt]
(1.06,\,1.02,\,1.01,\,0.43,\,0.26)
\quad\!\! &\mbox{[$SU(3)$, $\nS=3$].}
\end{array}
\right .
\phantom{x.x}
\label{eq:comparesfermions}
\eeq
\vspace{-0.6cm}

\noindent This shows that there is also a compression within the sfermion mass 
spectrum, as the ratio of the squarks to the slepton masses is decreased 
in the extended models compared to the MSSM in the $m_0^2 = 0$ limit, or 
more generally for any given value of $m_0$. This is because of the 
increased relative importance of the contribution to scalar masses from 
large renormalization scales where all of 
the gauge couplings and gaugino masses 
are larger.

Despite these compressions in the gaugino and sfermion sectors considered 
separately, the combined sparticle 
spectrum in the extended models is stretched 
rather than compressed compared to the MSSM. Comparing 
eqs.~(\ref{eq:comparegauginos}) and (\ref{eq:comparesfermions}), one 
observes that in each of the extended models, the bino is much lighter 
than the lightest slepton, and so the lightest supersymmetric particle 
(LSP) will be a neutralino. This is in
contrast to the 
well-known fact that the $m_0^2 = 0$ scenario in the MSSM 
problematically predicts a stau as the LSP. The 
supersymmetry-breaking flavor problem 
thus can be naturally solved by taking $m_0^2 \ll m_{1/2}^2$ in the 
extended models without running into the difficulties found in the 
gaugino mass dominated MSSM.

Now consider the soft supersymmetry breaking masses for the new particles. 
Assuming gaugino mass unification, the $G_X$ 
gaugino is heavier than all of the MSSM gauginos in the minimal $G_X = 
SU(2)$ and $SU(3)$ cases, but it is lighter than the MSSM gauginos if 
$G_X = SO(3)$. In terms of the unified gaugino mass parameter $m_{1/2}$, 
one finds at $Q = 1$ TeV:
\vspace{-0.5cm}
\beq
M_{\widetilde X} 
&=& m_{1/2}\! \times\! 
\left \{ \begin{array}{ll}
1.30
\quad &\mbox{[$SU(2)$, $\nS=0$],}
\\[-7pt]
0.051
\quad &\mbox{[$SO(3)$, $\nS=0$],}
\\[-7pt]
0.65
\quad\!\! &\mbox{[$SU(3)$, $\nS=3$].}
\end{array}
\right .
\eeq
\vspace{-0.5cm}

\noindent [Compare eq.~(\ref{eq:comparegauginos}).] 
The scalar members of the $D, \overline 
D$ and $L, \overline L$ multiplets get RG contributions to their soft 
masses from both $\GSM$ and $G_X$ gaugino loops. Therefore, they 
are heavier than their MSSM counterparts with the same gauge quantum 
numbers. For the case where $m_{1/2}$ dominates, one finds approximately
for the soft masses 
$m_{\widetilde D} = m_{\widetilde{\overline D}}$
and
$m_{\widetilde L} = m_{\widetilde{\overline L}}$, again at $Q = 1$ TeV:
\vspace{-0.5cm}
\beq
(m_{\widetilde D},\> m_{\widetilde L}) &=& m_{1/2} \times 
\left \{ \begin{array}{ll}
(1.97,\>1.40)
\quad &\mbox{[$SU(2)$, $\nS=0$],}
\\[-7pt]
(1.22,\>0.65) 
\quad &\mbox{[$SO(3)$, $\nS=0$],}
\\[-7pt]
(1.68,\>1.51) 
\quad &\mbox{[$SU(3)$, $\nS=3$].}
\end{array}
\right .
\label{eq:DLsoftmasses}
\eeq
\vspace{-0.5cm}

\noindent Also, for the minimal $SU(3)$ case with $\nS = 3$, one finds 
that $m_{\widetilde S} = m_{\widetilde {\overline S}} = 1.48 m_{1/2}$. 
This is only slightly lower than $m_{\tilde D}$ and $m_{\tilde L}$, 
because most of the RG contribution to these soft masses comes from 
$\widetilde X$ loops in this case, which are the same for all of the new scalars.

\begin{figure}[!tp]
\begin{center}
\vspace{-0.62cm}
\includegraphics[width=15.5cm,angle=0]{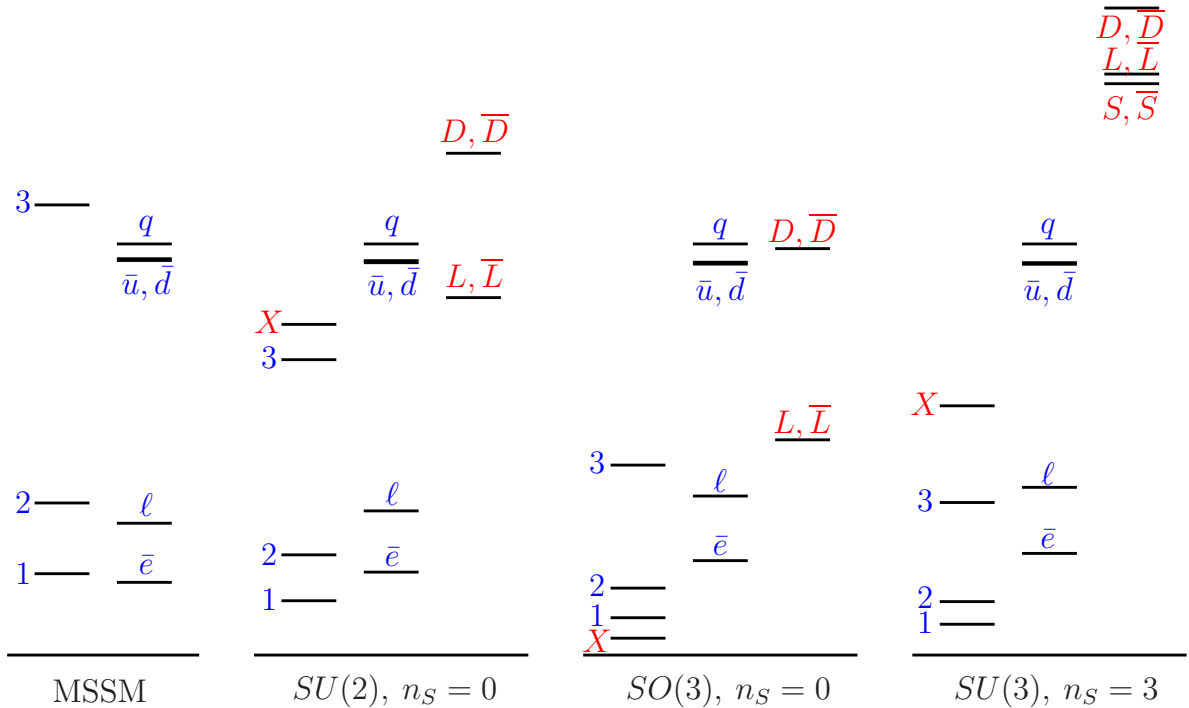}
\vspace{-0.4cm}
\end{center}
\caption{\label{fig:spectra}
Comparison of the soft supersymmetry-breaking mass spectra following from 
a unified input gaugino mass $m_{1/2}$ that dominates at the unification 
scale, for the MSSM and the minimal models with $G_X = SU(2), SO(3),$ and 
$SU(3)$. The labels $1,2,3,X$ refer to running gauginos masses for 
$U(1)_Y$, $SU(2)_L$, $SU(3)_c$ and $G_X$, respectively. The labels 
$q,\bar u, \bar d, \ell, \bar e$ refer to MSSM first family squark and 
slepton soft masses, and the soft scalar masses for vectorlike 
supermultiplets are labeled with symbols $D$, $\overline D$, $L$, 
$\overline L$, and (for the $G_X = SU(3)$ case with $\nS = 3$), $S, 
\overline S$. The value of $m_{1/2}$ is chosen so that the heaviest MSSM 
squark, $\tilde q$, has the same mass in each of the four cases.} 
\end{figure}

The qualitative features of the above results are illustrated in 
Figure \ref{fig:spectra}. The soft masses for the 
gauginos, the first-family sfermions, and the 
new scalars 
are shown for $Q = 1$ TeV. 
For purposes of comparison, $m_{1/2}$ is chosen 
so that the heaviest MSSM squark, $\tilde q$, has the same mass in each 
of the four cases. The mass spectra in the extended models are readily 
distinguishable from the usual ``mSUGRA" case parameterized by $m_{1/2}, 
m_0, A_0$. This is because obtaining such a large ratio of scalar masses 
to gaugino masses in mSUGRA, would require a large $m_0$, which in turn 
would lead to a much more compressed scalar mass spectrum. In contrast, 
the extended models are characterized by relatively heavy scalars which 
nevertheless maintain a significant hierarchy between squarks, 
left-handed sleptons, and right-handed sleptons, especially in the $G_X = 
SO(3)$ and $SU(3)$ cases.

In order to keep the discussion bounded, I will not give detailed results 
on the extended models with more singlets [$\nS>0$ for $G_X = SU(2)$ and 
$SO(3)$, and $\nS > 3$ for $G_X = SU(3)]$. However, the following 
qualitative features are notable. First, at one loop order, the presence of 
additional $\GSM$ singlets does not affect the RG running of MSSM-field 
soft terms, so the effects are rather mild on the gluino, wino, bino, and 
MSSM squark and slepton masses. Second, increasing $\nS$ will decrease both 
the $g_X$ gauge coupling and the $G_X$ gaugino soft mass at lower RG 
scales. Therefore, the $G_X$ gaugino mass 
$M_{\widetilde X}$ will be smaller compared to the 
MSSM gaugino masses $M_1$, $M_2$, and $M_3$ than in the cases shown in 
Figure \ref{fig:spectra}. Also, the soft masses $m_{\widetilde D} = 
m_{\widetilde{\overline D}}$ and $m_{\widetilde L} = 
m_{\widetilde{\overline L}}$ will become relatively smaller, tending 
towards the MSSM squark and slepton masses $m_{\tilde{\bar d}}$ and 
$m_{\tilde \ell}$ respectively. The soft masses for $m_{\widetilde S} = 
m_{\widetilde{\overline S}}$ will also decrease for larger $\nS$, 
although they are always heavier than $M_{\tilde X}$, which decreases 
faster for larger $\nS$. For $\nS > 0$, the Yukawa couplings $k,k'$ can 
also come into play, decreasing the $\widetilde{L}$, $\widetilde{\overline 
L}$, $\widetilde{S}$, and $\widetilde{\overline S}$ soft supersymmetry 
breaking masses.

\section{The $\mu$ parameter and the little hierarchy problem}
\label{sec:hierarchy}
\setcounter{equation}{0}
\setcounter{footnote}{1}

The supersymmetric little hierarchy problem is a subjective 
but inspirationally important puzzle 
which questions the naturalness of viable model parameters. The essence 
of it is that once one applies constraints from the non-observation of 
a Higgs boson and of superpartners at both LEP2 and the Tevatron,
the actual value of $m_Z$ might be considered surprisingly low 
for generic soft supersymmetry breaking parameters. 

The largest loop correction to the $h^0$ mass 
in the MSSM is given by (in the decoupling limit $m^2_{A^0} \gg 
m^2_{h^0}$):
\vspace{-0.4cm}
\beq
m^2_{h^0} &=& m_Z^2 \cos^2(2\beta) +
\frac{3 }{4 \pi^2} \sin^2\!\beta \>y_t^2 \biggl [
m_t^2 \, {\rm ln}\left (m_{\tilde t_1} m_{\tilde t_2} / m_t^2 \right )
+ c_{\tilde t}^2 s_{\tilde t}^2 (m_{\tilde t_2}^2 - m_{\tilde t_1}^2)
\, {\rm ln}(m_{\tilde t_2}^2/m_{\tilde t_1}^2) 
\nonumber
\\
&&+ c_{\tilde t}^4 s_{\tilde t}^4 \Bigl \lbrace
(m_{\tilde t_2}^2 - m_{\tilde t_1}^2)^2 - \frac{1}{2}
(m_{\tilde t_2}^4 - m_{\tilde t_1}^4)
\, {\rm ln}(m_{\tilde t_2}^2/m_{\tilde t_1}^2)
\Bigr \rbrace/m_t^2 \biggr ], 
\label{hradcorrmix}
\eeq
\vspace{-0.5cm}

\noindent 
where $m_{\tilde t_1}$ and $m_{\tilde t_2}$ are the top squark masses, 
and $c_{\tilde t}$ and $s_{\tilde t}$ are the cosine and sine of the 
top squark mixing angle $\theta_{\tilde t}$.
Now in the models discussed in \cite{Babu:2004xg} and this paper,
adding in the effects of the Yukawa coupling $k$ 
one finds the 
further estimated correction in the case $\mu_L \approx \mu_S \approx M_F$
with heavier scalars with masses of order $M_S$ \cite{Babu:2004xg}:
\vspace{-0.2cm}
\beq
\Delta m_{h^0}^2 = \frac{N}{4 \pi^2} k^4 v^2 \sin^4\beta \left [
f(x) + \frac{X_k^2}{M_S^2} (1 - \frac{1}{3x}) - \frac{X_k^4}{12 M_S^4} 
\right ] .
\label{eq:deltamhnew}
\eeq
\vspace{-0.5cm}

\noindent 
Here $x = M_S^2/M_F^2$ is the ratio of the average new scalar 
and
new 
fermion masses in the $L, \overline L, S, \overline S$ sector,
and $X_k = a_k/k - \mu \cot\beta$ is a mixing parameter for the scalars. 
The largest possible contributions 
come from the maximal (fixed-point) values of 
eq.~(\ref{eq:kfixed}).
As was pointed out in ref.~\cite{Babu:2004xg}, for $G_X = 
SU(3)$ eq.~(\ref{eq:deltamhnew}) 
is enough to raise the Higgs mass by tens of GeV, 
depending 
on the details of the fermion and scalar masses in the new sector.

From the point of view of the supersymmetric little hierarchy problem, 
even raising the 
Higgs mass by a few GeV is potentially helpful. However, one must also 
consider the effect of the new sector on the scalar potential.
The minimization of 
the Higgs potential in supersymmetry results in:
\vspace{-0.2cm}
\beq
m_Z^2 = -2 (|\mu|^2 + m^2_{H_u}) 
- \frac{1}{v_u} \frac{\partial}{\partial v_u} \Delta V
+ {\cal O}(1/\tan^2\beta) ,
\label{eq:mZSLH}
\eeq
\vspace{-0.5cm}

\noindent 
where $\Delta V$ is the radiative part of the effective potential, with 
$\tan\beta = v_u/v_d = \langle H_u^0 \rangle/\langle H_d^0 \rangle$ and  
$v_u$ treated as a real variable in the partial differentiation. In 
general, without further theoretical structure, 
$\mu$ and $m^2_{H_u}$ have no reason 
to be related, since $\mu$ is a supersymmetry-preserving parameter and 
$m^2_{H_u}$ is supersymmetry-breaking. In the MSSM with generic 
parameters,
one finds that $-m^2_{H_u}$ 
tends to be  
much larger than $m_Z^2$, and eq.~(\ref{eq:mZSLH}) 
seems to imply a percent-level fine-tuning of the difference between 
$|\mu|^2$ and $m_{H_u}^2$. 

It is not 
possible to rigorously quantify fine tuning, since there can be no such 
thing as an objective measure on parameter space. 
Nevertheless, qualitative trends can be identified, and
an obvious approach is to consider models with smaller
predicted values of $-m_{H_u}^2$ at the weak scale to be more likely 
than those with very large $-m_{H_u}^2$,
because then the fractional tuning required between it and $|\mu|^2$ will 
be less. This in turn means that smaller values of $|\mu|$ are more 
likely than very large values, since this is determined by 
eq.~(\ref{eq:mZSLH}). 

With this in mind, it is interesting to consider how the MSSM and its 
extensions, and variations of the most popular models of supersymmetry 
breaking, affect the weak-scale predictions for $-m^2_{H_u}$. For 
example, in the MSSM with $\tan\beta=10$ and $m_t = 173.3$ GeV, one finds 
from RG running at $Q = 1$ TeV in terms of the GUT-scale 
input parameters $m_{1/2}$, $A_0$ and $m_0$:
\vspace{-0.2cm}
\beq
-m_{H_u}^2 = 1.65 m_{1/2}^2 - 0.40 m_{1/2} A_0 + 0.11 A_0^2 
-0.022 m^2_0 .
\eeq
\vspace{-0.6cm}

\noindent 
This formula shows that $-m_{H_u}^2$, and therefore $|\mu|^2$, and 
therefore the level of fine-tuning required, increase with the 
gaugino squared masses. 
In extended models the gaugino masses at the unification scale 
have a varying relationship with the gaugino masses at the weak scale, 
which are more closely related to the physical
masses, so it is useful to reformulate this in terms of the running gluino 
mass parameter $M_3$ also evaluated at $Q = 1$ TeV:
\vspace{-0.2cm}
\beq
-m_{H_u}^2 = 0.32 M_3^2 - 0.18 M_3 A_0 + 0.11 A_0^2 -0.022 m^2_0,
\qquad \mbox{[MSSM]}
.
\label{eq:m2HuMSSM}
\eeq
\vspace{-0.7cm}

\noindent 
In fact, most of the dependence on the gaugino masses comes from the 
gluino mass
\cite{Kane:1998im}, so this formula is approximately valid even for moderate
deviations from gaugino mass universality. One can note that for a gluino 
mass of order 500 GeV, and small $|A_0|$, $-m_{H_u}^2$ is only of 
order 
(280 GeV)$^2$, so that the tuning needed to get $m_Z^2$ in 
eq.~(\ref{eq:mZSLH})
is of order 5\%. The problem is that (although there is still 
considerable variation among models, particularly for large $|A_0|$) 
lower values of $M_3$ typically give a prediction 
for $m_{h^0}$ that is smaller than 114 GeV, and higher values of $M_3$ 
require even more delicate cancellation
between $-m_{H_u}^2$ and $|\mu|^2$.
The ``focus point" region 
\cite{Chan:1997bi,Feng:1999mn}
occurs due to the small negative coefficient of 
$m_0^2$ in eq.~(\ref{eq:m2HuMSSM}), which allows a cancellation between
the $M_3^2$ and $m_0^2$ terms for very large $m_0^2$, leading to a small 
value of $-m_{H_u}^2$ and therefore small $|\mu|^2$. However, this also 
can be judged to be fine-tuned, as the large value of $m_0^2 $ has to be 
finely adjusted, 
given a value of the ostensibly independent parameter $M_3$.

We can now compare with the situation for the models in the present 
paper.
In the minimal $SU(2)_X$ model with $\nS = 0$, one finds instead from 
the RG running:
\vspace{-0.2cm}
\beq
-m_{H_u}^2 = 1.07 M_3^2 - 0.44 M_3 A_0 + 0.12 A_0^2 -0.16 m^2_0,
\qquad [SU(2)_X,\> \nS = 0].
\label{eq:m2HuSU20}
\eeq
\vspace{-0.7cm}

\noindent 
again for $-m_{H_u}^2$ and $M_3$ evaluated at $Q = 1$ TeV,
assuming gaugino and scalar mass universality, $\tan\beta=10$ and $m_t = 
173.3$ GeV. (The 
coefficients change, but not very radically, for 
larger $\nS$.) The larger coefficient of $M_3^2$ indicates that this is 
naively even more fine-tuned than the MSSM. However, this effect is not 
without compensation; as Figure \ref{fig:spectra} shows, one does not 
need 
as large a gluino mass to get large squark masses, which in turn lead to
large positive contributions 
to $m_{h^0}^2$ from eq.~(\ref{hradcorrmix}). Also,
the larger negative coefficient of $m_0^2$ means that the analog of the
MSSM focus point region occurs at much smaller values 
of $m_0^2$ in this extended model. Since 
there is no such thing as an objective quantitative measure of 
fine-tuning, I choose not to attempt to make a definitive statement 
beyond observing the competing factors just mentioned.

For $SO(3)_X$ with $\nS = 0$, the analogous formula becomes:
\vspace{-0.2cm}
\beq
-m_{H_u}^2 = 3.36 M_3^2 - 0.91 M_3 A_0 + 0.13 A_0^2 -0.26 m^2_0,
\qquad [SO(3)_X,\> \nS = 0] .
\eeq
\vspace{-0.7cm}

\noindent 
Similarly, for the minimal viable $SU(3)_X$ model with $\nS = 3$ and 
small Yukawa 
couplings $k,k'\approx 0$, the analogous formula becomes:
\vspace{-0.2cm}
\beq
-m_{H_u}^2 = 5.97 M_3^2 - 1.28 M_3 A_0 + 0.13 A_0^2 -0.28 m^2_0,
\qquad [SU(3)_X,\> \nS = 3] .
\eeq
\vspace{-0.7cm}

\noindent 
In both of these cases, the situation again seems subjectively worse with 
respect to fine-tuning than the MSSM, 
due to the much larger coefficient of $M_3^2$. 
As shown in Figure \ref{fig:spectra}, one does naturally get much 
larger squark masses for a given $M_3$, again leading to larger radiative 
corrections to $m_{h^0}^2$. However, with these large coefficients to $M_3$, 
the direct (but model dependent) constraints on the gluino mass from 
Tevatron come into play in a significant way. Again, there is a chance 
for more cancellation between the gaugino and scalar contributions due to 
the negative coefficient of $m_0^2$.

It is also interesting to consider the situation for $\nS \geq 1$ with 
large Yukawa couplings near the fixed points of eq.~(\ref{eq:kfixed}). 
For $SU(2)_X$ with $\nS 
= 1$ and the fixed-point value $k = k_{\rm fixed} = 0.88$ at $Q = 1$ 
TeV, one finds:
\vspace{-0.2cm}
\beq
-m_{H_u}^2 = 1.28 M_3^2 - 0.15 M_3 A_0 + 0.04 A_0^2 + 0.72 m^2_0,
\qquad [SU(2)_X,\> \nS = 1,\> k = k_{\rm fixed}] .
\eeq
\vspace{-0.7cm}

\noindent 
Here the coefficient of the gaugino mass squared is even 
larger than for $k=0$ [compare eq.~(\ref{eq:m2HuSU20})], 
and the coefficient of the scalar squared mass
$m_0^2$ is large and positive, eliminating the possibility of 
cancellation to achieve a smaller $-m_{H_u}^2$. Similar 
results obtain 
for $SO(3)_X$ with $\nS = 1$ and $k = k_{\rm fixed} = 0.76$ at $Q = 1$
TeV:
\vspace{-0.2cm}
\beq
-m_{H_u}^2 = 3.46 M_3^2 - 0.27 M_3 A_0 + 0.05 A_0^2 + 0.89 m^2_0,
\qquad [SO(3)_X,\> \nS = 1,\> k = k_{\rm fixed}] ,\phantom{xx}
\eeq
\vspace{-0.7cm}

\noindent 
and for $SU(3)_X$ with $\nS = 3$ and one 
$k = k_{\rm fixed} = 1.32$ at $Q = 1$ TeV:
\vspace{-0.2cm}
\beq
-m_{H_u}^2 = 17.0 M_3^2 - 0.27 M_3 A_0 + 0.02 A_0^2 + 0.71 m^2_0,
\qquad [SU(3)_X,\> \nS = 3,\> k = k_{\rm fixed}] .\phantom{xx}
\eeq
\vspace{-0.7cm}

\noindent 
Therefore, even though the fixed-point Yukawa coupling can give large 
positive contributions to $m_{h^0}^2$, there is a quite detrimental 
effect on the fine-tuning needed to obtain the observed $m_Z$ in models 
that have heavy enough gluinos (and charginos) to have evaded discovery 
at the Tevatron and LEP2. Similar effects have been noted before in the 
case of vector-like fermions without an additional gauge group in 
refs.~\cite{Babu:2008ge}, \cite{Martin:2009bg}, \cite{Martin:2010dc}.

Qualitatively, the model with $SU(2)_X$ and no new Yukawa coupling seems 
to be the least fine-tuned of the extended models. Adding new Yukawa 
couplings, despite increasing $m_{h^0}$, does not clearly alleviate the 
little hierarchy problem, and arguably make it much worse, especially in 
the cases of $SO(3)_X$ and $SU(3)_X$.

\section{Collider phenomenology of the quirks}
\label{sec:pheno}
\setcounter{equation}{0}
\setcounter{footnote}{1}

In this section, I will consider some features of the phenomenology of 
the quirks in the models discussed above, following for the most part 
general ideas and results from refs.~\cite{Kang:2008ea}, 
\cite{Burdman:2008ek}, \cite{Cheung:2008ke}, and \cite{Harnik:2008ax}. 
For simplicity, I will consider only the fermions from the $D, \overline 
D$, $L, \overline L$, multiplets, and not their scalar partners. This is 
because supersymmetry breaking effects provide for the scalars 
(``squirks") and the $G_X$ gaugino to have much larger masses, making 
them less immediately relevant for collider searches. (Even if they had 
the same masses, squirks would have much smaller production 
cross-sections than fermionic quirks. The $G_X$ gauginos will not be 
produced directly in tree-level processes at colliders at all.) When 
produced, the squirks will decay promptly to quirks and MSSM gauginos. 
The $G_X$ gaugino can undergo a three-body decay to a quirk, antiquirk 
and MSSM gaugino, if kinematically allowed. In this section, I will use 
the same symbols for the fermions as for the chiral supermultiplets to 
which they belong.

I will also assume for simplicity, and motivated by the results of 
the previous section, that the mixing of the $\GSM$ singlets 
$S, \overline S$ with the doublets $L, \overline L$ due to the Yukawa 
couplings $k,k'$ is small in most of the following. This implies that $S, 
\overline S$ decouple from collider phenomenology. The charged $(L^-, 
\overline L^+)$ and neutral $(L^0, \overline L^0)$ fermions form two 
Dirac fermion-antifermion pairs, each with tree-level mass $\mu_L$. 
However, radiative 
corrections split the masses slightly, with $\Delta m \equiv m_{L^-} 
- m_{L^0}$ always positive. (If present, the Yukawa interactions 
$k,k'$ that cause mixing with $S, \overline S$ would increase this 
splitting, so the lightest non-colored fermion is always neutral.) 
One finds $\Delta m > 270$ MeV for $\mu_L > 100$ GeV, 
with $\Delta m$ 
approaching 355 MeV asymptotically for large $\mu_L$ 
\cite{Thomas:1998wy}. This means that the decays
\vspace{-0.2cm}
\beq
\overline L^+ &\rightarrow& \overline L^0 \ell^+ \nu_\ell,
\>\>\overline L^0 \pi^+, 
\\
L^- &\rightarrow& L^0 \ell^- \overline \nu_\ell, \>\> L^0 \pi^- ,
\eeq 
\vspace{-0.7cm}

\noindent 
(and decays to more pions or other SM hadrons if non-zero $k,k'$ 
increase $\Delta m$) mediated by the $W$ boson are always kinematically 
allowed, and will occur with decay lengths of order centimeters 
\cite{Thomas:1998wy} due to the small available kinematic phase space. 
Therefore, the lifetime of $\overline L^+, L^-$ is large compared to 
other processes to be discussed below; in particular they will form 
quirk-antiquirk bound states and annihilate before they decay. In the 
simplest scenario (barring additional couplings to be described in the 
next paragraph), the neutral quirk Dirac fermions $L^0, \overline 
L^0$ are completely stable, as are the colored quirks $D, 
\overline D$ with charges $\pm 1/3$. 
Note that none of the quirks can mix with the 
Standard Model fermions because of $G_X$ conservation, 
so the lightest quirk is always stable. Such stable 
fermions could present a challenge for the standard cosmology with a high 
reheat temperature but need not be a disaster 
\cite{Kang:2006yd},\cite{Jacoby:2007nw},\cite{Kang:2008ea}.

If $\nS \geq 1$ and the pairs $L, \overline L$ and $S,\overline S$ have 
the opposite matter parity from each other, then the Yukawa couplings 
$k,k'$ are forbidden, but the superpotential term
\vspace{-0.2cm}
\beq
W = \lambda_\ell \overline S\hspace{0.015in} \overline L \hspace{0.005in} \ell
\eeq
\vspace{-0.7cm}

\noindent 
is allowed, with $\ell$ an MSSM $SU(2)_L$ doublet lepton. This provides 
additional possible decay modes $\overline L^+ \rightarrow 
S \tilde \ell_L^+$ and $L^- \rightarrow 
\overline S \tilde \ell^-_L$, if 
$|\mu_L| > |\mu_S| + m_{\tilde \ell_L}$ so that
these decays are kinematically allowed, or alternatively 
with the sleptons off-shell. These decays are not automatically kinematically 
suppressed, and so could happen promptly before quirk-antiquirk
annihilation occurs.
The fermions $D$ and ${\overline D}$ 
may also have additional decay modes, if the possible
superpotential terms
\vspace{-0.2cm}
\beq
W = 
\lambda_{q} L \overline D q + 
\lambda_{\overline d} \overline S D \overline d , 
\label{eq:Ddecay}
\eeq
\vspace{-0.7cm}

\noindent 
are present.
The first of these terms
is only allowed if $G_X$ is either $SU(2)$ or $SO(3)$, and if
$L, \overline L$ 
have the opposite matter parity of $D, \overline D$ (if matter parity is conserved). 
It permits squark exchange to mediate the decays
\vspace{-0.18in}
\beq
&&D \rightarrow 
L^0 \tilde d_L,
\quad
L^- \tilde u_L,
\label{eq:psiDdecaysA}
\\
&&\overline{D} \rightarrow \overline L^0 \tilde d_L^*,
\quad
\overline L^+ \tilde u_L^*
,
\label{eq:psiDdecaysB}
\eeq 
\vspace{-0.33in}

\noindent with the MSSM squarks possibly off-shell due to kinematics.
The second term in eq.~(\ref{eq:Ddecay}) is only allowed if
$\nS\geq 1$ and $S,\overline S$ have the opposite matter parity of
$D, \overline D$. 
Then MSSM squark exchange can mediate the decays 
\beq
&&
D \rightarrow S \tilde d_R,\qquad
\overline{D} \rightarrow \overline S \tilde d_R^*
,
\label{eq:psiDdecaysC}
\eeq 
again with the squarks possibly off-shell.
Whether these decays can be important depends on the kinematics as 
well as the size of 
$\lambda_\ell$, $\lambda_q$, and $\lambda_{\overline d}$. For simplicity, 
they will be assumed to be absent or at least too small to make a 
difference below, except where noted otherwise.
 
The important direct pair-production processes for the new fermions are
\vspace{-0.1in}
\beq
pp &\rightarrow& D {\overline D},
\\
pp&\rightarrow & Z^{(*)}, \gamma^{(*)} \>\rightarrow\> \overline L^+ 
L^-,\>\,
\overline L^0 L^0 ,
\\
pp&\rightarrow & W^{+(*)} \> \rightarrow \> \overline L^+ L^0  ,
\label{eq:proplus}
\\
pp&\rightarrow & W^{-(*)} \> \rightarrow \> \overline L^0 L^- ,
\label{eq:prominus}
\eeq
for the LHC, with the obvious substitution of $p\overline p$  for 
the Tevatron. Pair-produced quirks with masses much larger than $\Lambda$ 
will move apart from each other with typically semi-relativistic speeds, 
and as described in \cite{Kang:2008ea}, 
will be connected by $G_X$ flux strings with tension $\sigma$. From the 
lattice, there is an estimate (see Table 7 and eq.~(11) of 
\cite{Allton:2008ty}):
\vspace{-0.25in}
\beq
\sqrt{\sigma} = \Lambda_{\MSbar} \times 
\biggl \{ \begin{array}{l}
1.73\quad SU(2)_X,
\\[-6pt]
1.86\quad SU(3)_X,
\end{array}
\Biggr.
\eeq
\vspace{-0.3in}

\noindent
so that the maximum string length in a given hard scattering event with
kinetic energy $\Delta E$ in the center-of-momentum frame is 
\beq
L = \frac{\Delta E}{\sigma} \approx 
6 \, \mbox{mm}\> 
\left (\frac{\Delta E}{\mbox{100 GeV}}\right )
\left (\frac{\mbox{keV}}{\Lambda_{\MSbar}}\right )^2 .
\eeq
Therefore, the lengths of such strings, although much larger than 
$\Lambda^{-1}$, 
will typically be less than 1 mm 
for $\Lambda$ greater than a few  keV. 
From Table \ref{tab:Lambda}, one finds that the quirky flux 
strings will be microscopic for $SU(2)_X$ with $\nS \leq 3$ and for 
$SU(3)_X$ with $\nS \leq 9$, assuming that $g_X$ is unified with the SM 
gauge couplings and all singlets charged under $SU(N)_X$ are heavier than 
$\Lambda$. 

For the case of $G_X = SO(3)$, the situation is quite different, because 
from Table \ref{tab:Lambda} the confinement distance scale $\Lambda^{-1}$ 
is literally astronomical, at least of order 
the Earth's orbit around the Sun 
even in the minimal model. The quirks in this 
case are essentially free particles with multiplicity 3 times larger than 
expected from their SM quantum numbers. Note that even if one rejected 
the unification of $g_X$ with the SM gauge couplings in this model to 
arrive at a much larger $\Lambda$, the fact that the supermultiplets are 
in the adjoint representation of the Lie algebra means that they would 
not form stable flux tubes of the type discussed in \cite{Kang:2008ea} 
when pair-produced, even if any $S,\overline S$ fields are heavier than 
$\Lambda$ (so $n=0$). Instead, pair-produced particles charged under 
$SO(3)_X$ would each bind to a gauge boson to form two stable 
$G_X$-singlet states with size of order $\Lambda^{-1}$, allowing the flux 
tube to break. Although the new fermions behave like free stable 
particles when pair-produced at colliders, the fact that they will come 
in three-fold exactly degenerate multiplets will in principle allow a 
determination of their nature from their production cross-sections. In 
the simplest case, $D, \overline D$ and $\overline L^0, L^0$ 
will be absolutely stable, with $\overline L^+, L^-$ having decays 
to $\overline L^0, L^0$ via soft pion or lepton emission as 
discussed above. The $\overline L^0, L^0$ are only weakly 
interacting and thus invisible, but known collider search strategies 
\cite{Abe:1989es}
for stable strongly interacting particles apply for $D, \overline D$. 
However, as 
noted above, $D, \overline D$ may be able to promptly decay according to 
eqs.~(\ref{eq:psiDdecaysA})-(\ref{eq:psiDdecaysC}), depending on both 
kinematics and the allowed superpotential terms. 
If so, then the signatures will always contain $\ETmiss$, and 
will resemble those for ordinary MSSM squarks.

For the remainder of this section, consider the cases of 
$SU(2)_X$ and $SU(3)_X$, with the $G_X$ confinement scale less than the 
masses of the quirks that have SM gauge interactions, and stable 
microscopic flux strings joining the quirk-antiquirk pairs. The 
quirk-antiquirk pair will then form an exotic bound state with invariant 
mass given approximately by the total center-of-momentum energy of the 
hard partonic scattering that produced them. This quirk-antiquirk string 
state can lose energy either by $G_X$-glueball emission, by radiation of 
many soft photons, or in the case of the $D\overline D$ 
state by radiation of numerous soft pions, 
a ``hadronic fireball" \cite{Kang:2008ea}. The 
large multiplicity of soft pions or photons may be detectable as 
anomalous ``underlying events" \cite{Kang:2008ea,Harnik:2008ax} that 
accompany the hard scattering production, and may be used as an 
additional tag to dramatically reduce backgrounds.

If the quirk and antiquirk lose most of their initial relative kinetic 
energy before annihilating, they will briefly form a ``quirkonium" bound 
state which then decays to two or three hard partons with invariant mass 
peaked at twice the mass of the quirk \cite{Kang:2008ea}.
Alternatively, however, 
the neutral and colorless quirk and antiquirk states might  
\cite{Kang:2008ea} have a prompt 
annihilation before they can lose enough energy to form 
a low-lying quirkonium state. 
In that case, the final states will have a 
broad distribution of annihilation products, which will therefore be much 
harder to discern above hadron collider backgrounds. It is difficult 
to estimate in advance what proportion of the events will fall into these two 
categories, due to the non-perturbative nature of the energy loss 
mechanisms, which do not have direct analogs in experimentally known 
hadronic physics.

For the weakly interacting quirks, and for the strongly interacting 
quirks if $\Lambda > \Lambda_{\rm QCD}$, one might suspect the 
non-perturbative interactions by which the quirk-antiquirk string state 
loses energy to be dominated by $G_X$-glueball emission. However, this is 
quite uncertain, and can be suppressed or even eliminated by 
kinematics if $\Lambda \gg \Lambda_{\rm QCD}$. 
The masses of the $G_X$ glueballs have been 
estimated by lattice computations 
\cite{Teper:1998kw}-\cite{Lucini:2001ej}, \cite{Allton:2008ty},
with the results for the 
lightest two glueball states with $J^{PC} = 0^{++}$ and $2^{++}$:
\beq
m_{0^{++}} &=& 6.7 \Lambda_{\MSbar},
\\
m_{2^{++}} &=& 9.6 \Lambda_{\MSbar} .
\eeq
There are other heavier glueball states $0^{++*}$, $3^{++}$, $0^{-+}$,
$2^{-+}$, $0^{-+*}$, $2^{-+*}$, and, for $SU(3)_X$ only there are also
states with odd $C$,
$1^{+-}$, $3^{+-}$, $2^{+-}$, $0^{+-}$, $1^{--}$, $2^{--}$ $3^{--}$, with
masses ranging up to about $3 m_{0^{++}}$. 
As can be seen from Tables \ref{tab:Lambda} and \ref{tab:Lambdalow}, in 
the case of $SU(3)_X$ these glueballs should have 
masses in the hundreds of GeV range for the minimal case of $\nS = 3$ 
and so could be comparable 
in mass or even heavier than the lighter quirks, and should be in 
the tens of GeV range for $\nS = 4$. This would prohibit 
energy loss of the quirk-antiquirk flux string states into $G_X$-glueballs. 
For the other cases listed in Table \ref{tab:Lambda}, decays of the flux 
strings to $G_X$-glueballs should be allowed, but perhaps kinematically 
suppressed, leading 
to considerable uncertainty in the number of $G_X$-glueball states 
emitted and the likelihood of the quirk-antiquirk string state to lose 
most of its energy before annihilating. It is also possible that a few 
$G_X$-glueballs will be produced in the original hard scattering 
production.

If produced, the detection of $G_X$-glueballs is problematic.
Their decay widths can be estimated for 
$SU(3)_X$ using 
eqs.~(17), (23) and (30) of 
ref.~\cite{Juknevich:2009ji} (see also ref.~\cite{Juknevich:2009gg})
with matrix elements from eqs.~(38) and (62) of 
ref.~\cite{Chen:2005mg}:
\beq
\Gamma(0^{++} \rightarrow gg) &=& 360\, \alpha_S^2 
{\Lambda^9}/{\mu_D^8},
\\
\Gamma(2^{++} \rightarrow gg) &=& 0.12\, \alpha_S^2 
{\Lambda^9}/{\mu_D^8},
\\
\Gamma(0^{-+} \rightarrow gg) &=& 24\, \alpha_S^2 
{\Lambda^9}/{\mu_D^8}
.
\eeq
The results for $SU(2)_X$ should be comparable and slightly smaller. This
leads to proper decay lengths for $G_X$-glueballs of order
\beq
c\tau = \left (\frac{0.2}{\alpha_S}\right )^2
\left (\frac{\mu_D}{\mbox{100 GeV}}\right )^8
\left (\frac{\mbox{GeV}}{\Lambda}\right )^9 \times
\left \{ \begin{array}{l}
0.14\>{\rm meters}\quad \mbox{(for $0^{++}$)},
\\[-9pt]
400\>{\rm meters}\quad \mbox{(for $2^{++}$)},
\\[-9pt]
2\>{\rm meters}\quad \mbox{(for $0^{-+}$)} 
.
\end{array}
\right.
\eeq
If $\Lambda \approx 1$ GeV as expected for the minimal $SU(2)_X$ model, a 
sizable fraction of the $0^{++}$ decays might occur within the detector, but 
only if $\mu_D$ is less than roughly 150 GeV, which may be
ruled out already by Tevatron data (see below). For larger $\mu_D$ or 
smaller $\Lambda$, the decays of the $G_X$-glueball will occur outside of 
the detector and will be invisible. For much larger $\Lambda$ as occurs 
in the $SU(3)_X$ model with $\nS = 3$ or 4, the decays may occur within 
the detector for any $\mu_D$, but then the production of $G_X$ glueballs 
in the flux-tube energy loss processes 
is likely irrelevant anyway due to kinematic suppression or prohibition. 
Even if the $G_X$ glueballs are produced and decay promptly, the 
main decay is likely to a pair of gluons, and the resulting dijet mass 
peak signal from these decays will have to compete with a huge background 
from QCD. To have a significant branching fraction to $\gamma\gamma$, which has much smaller backgrounds, one
can take $\mu_L < \mu_D$, with a leading-order estimate 
\cite{Juknevich:2009ji,Juknevich:2009gg}:
\beq
\frac{{\rm BR}(0^{++} \rightarrow \gamma\gamma)}{
      {\rm BR}(0^{++} \rightarrow gg)}
= \frac{8 \alpha^2}{9 \alpha_S^2} \left (\frac{1}{4} + \frac{3 \mu_D^4}{4 
\mu_L^4} 
\right )^2  
.
\eeq
For example, with $\alpha_S \approx 0.2$, this ratio is of order 0.001 
for $\mu_L = \mu_D$, but it
rises to about 0.18 if $\mu_L/\mu_D = 0.5$, and is greater than 1 if
$\mu_L/\mu_D$ is less than $0.4$. The 
non-perturbative 
nature of the $G_X$ glueball production mechanisms means that 
the diphoton signal strengths, if any, are extremely 
difficult to estimate even roughly, but to have even a hope of observation 
would seem to 
require $\mu_L < \mu_D$ and $\Lambda$ of order a few GeV (not too small 
for $c\tau$ to be large, but not too large for $G_X$-glueball production to be 
kinematically suppressed). Nevertheless, 
given the uncertainties involved, this possibility 
highlights the general importance of searching for narrow diphoton peaks 
at large invariant masses at the LHC; this type of signal could also 
arise not only for the classic diphoton signal for a low-mass Higgs 
scalar boson, but also for stoponium \cite{stoponium1,stoponium2} or for 
Kaluza-Klein gravitons in theories with low-scale gravity 
\cite{Lemaire:2006kf}.

Probably the most optimistic scenario for detecting the quirks occurs in 
the case that $D\overline D$ are strongly produced at a hadron 
collider and manage to lose most of their initial relative kinetic energy 
stored in the $G_X$ flux tube by radiating soft pions and/or $G_X$ 
glueballs, arriving at a low-lying quirkonium state with mass $\approx 
2\mu_D$ before finally annihilating in a color-singlet 
$S$-wave ${}^{2S+1}L_J = {}^1S_0$ 
($\eta$) or ${}^3S_1$ ($\psi$) state. The most promising channel for 
detecting the quirkonium peak is $\mu^+\mu^-$. The relevant annihilation 
decay widths for a ${}^3S_1$ state can be inferred from 
refs.~\cite{Barger:1987xg} and \cite{Cheung:2008ke}:
\beq
&&\Gamma (\psi \rightarrow f \overline f) = 4 \alpha^2 
e_D^2 
N_c^f \beta_f
\biggl [ (1 + 2 R_f) 
\Bigl (e_f - \frac{g_f^V}{c_W^2 (1-R_Z)} \Bigr )^2 + 
\Bigl (\frac{\beta_f g_f^A}{c_W^2 (1-R_Z)} \Bigr )^2 
\biggr ] \Gamma_0,\phantom{xxxxx}
\label{eq:DDffdecay}
\\
&&\Gamma (\psi \rightarrow W^+W^-) =
\frac{\alpha^2 e_D^2 \beta_W^3}{4 c_W^4} 
\frac{1 + 20 R_W + 12 R_W^2}{(1 - R_Z)^2}
\Gamma_0
\\
&&\Gamma (\psi \rightarrow ggg) = 
\frac{40 \alpha_S^3}{81 \pi} (\pi^2 - 9) \Gamma_0
\label{eq:DDgggdecay}
\\
&&\Gamma (\psi \rightarrow XXX) = \frac{\alpha_X^3}{3 \pi} 
\frac{(N^2 - 1)(N^2 - 4)}{N^3} (\pi^2 - 9) \Gamma_0
\eeq
where $e_f = (2/3, -1/3, -1, 0)$ and $N_c^f = (3,3,1,1)$ and $g_f^A = 
(1/4, -1/4, -1/4, 1/4)$ for $f = (u,d,e,\nu)$ respectively, and $g_f^V = 
g_f^A - e_f s_W^2$, and $s_W$ and $c_W$ are the sine and 
cosine of the weak mixing angle, and $e_D = -1/3$, and $R_i = m_i^2/M^2$, 
where $M\approx 
2\mu_D$ is the quirkonium mass, and $\beta_i = \sqrt{1 - 4 R_i^2}$, and 
$\Gamma_0$ is a common normalization proportional to the square 
of the wavefunction at the origin. The $G_X$ gluon is represented by 
$X$. Note that final states $gg$, $ZZ$, 
$Z\gamma$, $\gamma\gamma$ and $XX$ do not occur in $\psi$ decays. The 
final states $Zgg$, $\gamma XX$, $ZXX$ do occur, but with 
branching ratios that turn out to be very small. For $N=2$, the decay to 
three $SU(2)_X$ gauge bosons vanishes due to the $N^2-4$ factor, and for
$SU(3)_X$ with 
$N=3$ the $XXX$ decay
will be kinematically forbidden or at least highly 
suppressed by the large masses of the $G_X$ glueballs that would have to
be the final result of $G_X$-hadronization. Therefore 
final states involving $G_X$ glueballs should not play a significant 
role in quirkonium decays.

For ${}^1S_0$ states, the dominant decay is to $gg$ or $XX$, 
and $f \overline f$ does 
not occur at all at leading order. If we assume that the spin state is 
randomized by the non-perturbative processes that lose the initial 
relative kinetic energy, so that $\psi$ and $\eta$ states are populated 
in the ratio of 3 to 1, then the branching ratio of quirkonium to 
leptons 
should be given by 3/4 of the branching ratio indicated by 
eqs.~(\ref{eq:DDffdecay})-(\ref{eq:DDgggdecay}). Numerically this 
yields\footnote{The estimate in section 5.6 of 
\cite{Kang:2008ea} is parametrically different, and numerically 
smaller by a factor $\sim$5.} 
BR$(D\overline D \rightarrow 
\mu^+\mu^-) = 0.093$ very nearly independent of the mass. 
It should be noted that this branching ratio does 
not apply to prompt annihilation of the quirks before they have settled 
into a color- and $G_X$-singlet quirkonium state; that branching ratio 
will be much smaller, and will not lead to a sharp dimuon peak, and so 
leads to a more pessimistic case.

\begin{figure}[!tp]
\begin{minipage}[]{0.52\linewidth}
\begin{flushleft}
\hspace{0.253cm}\includegraphics[width=7.52cm,angle=0]{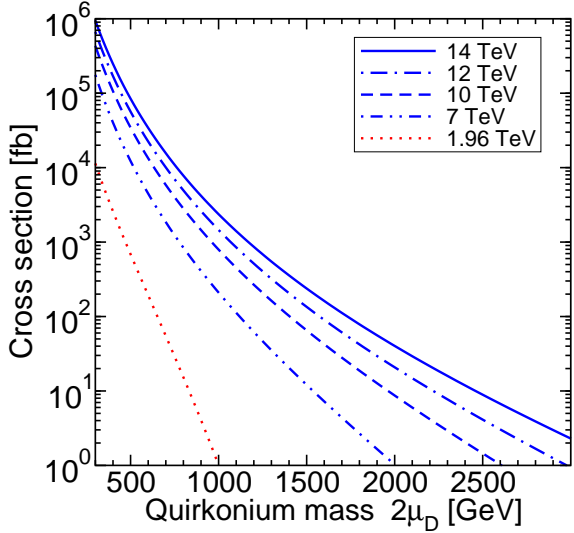}
\end{flushleft}
\end{minipage}
\begin{minipage}[]{0.47\linewidth}
\caption{\label{fig:Dquirkprod} The production 
cross-section for $D \overline D$ in $pp$ collisions at $\sqrt{s} = 14$, 
12, 10, and 7 TeV, and in $p\overline p$ collisions at 
$\sqrt{s} = 1.96$ TeV, obtained at leading order using 
CTEQ5LO \cite{CTEQ5} parton distribution functions (PDFs) with $Q=\mu_D$.}
\end{minipage}
\end{figure}

In the most optimistic case that most of the $D \overline D$ 
states annihilate after losing most of their excess energy, 
there are good prospects for detection at hadron colliders, because the 
signal production is strong and peaked in invariant mass, while 
the dominant background is electroweak (Drell-Yan) and diffuse. The total 
production cross section at the 
Tevatron and at various LHC energies is shown in Figure
\ref{fig:Dquirkprod}. The CDF collaboration has published 
\cite{Aaltonen:2008ah} a limit on cross-section times branching ratio for 
new states that decay to $\mu^+\mu^-$, based on 2.3 fb$^{-1}$ of 
$p\overline p$ collisions at the Tevatron. Comparing the relevant spin-0 
limit from Figure 3 in \cite{Aaltonen:2008ah} to the results shown in 
Figure \ref{fig:Dquirkprod} of the present paper and using the estimate
BR$(\mu^+\mu^-) = 0.093$ from above, I obtain the lower mass 
bound $\mu_D > 375$ GeV in this optimistic case.
 
\begin{figure}[!tp]
\begin{minipage}[]{0.52\linewidth}
\begin{flushleft}
\vspace{-0.3cm}
\includegraphics[width=7.8cm,angle=0]{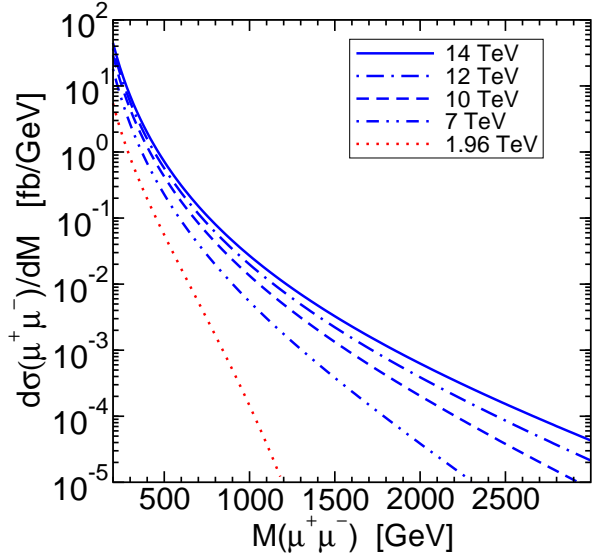}
\end{flushleft}
\end{minipage}
\begin{minipage}[]{0.47\linewidth}
\caption{\label{fig:DYmumu} The leading-order 
differential production cross-section for the $\mu^+\mu^-$ background, 
through $\gamma^*$ and $Z^*$-mediated (Drell-Yan) processes
in $pp$ collisions at $\sqrt{s} = 14$, 12, 10, and 7 TeV, and in
$p\overline p$ collisions at $\sqrt{s} = 1.96$ TeV, as a function of the 
invariant mass of the $\mu^+\mu^-$ pair. 
CTEQ5LO \cite{CTEQ5} PDFs were used with $Q=\sqrt{\hat s}$.}
\end{minipage}
\end{figure}

At the LHC, the invariant mass resolution for high-mass dimuons should be 
of the order of 5\% \cite{Belotelov:2006bb} for the CMS detector. 
Therefore, as a rough estimate of the discovery reach, I consider a mass 
window from $0.9M$ to $1.1M$ where $M \approx 2\mu_D$ is the quirkonium 
mass, and require that $S/\sqrt{B}$ exceeds 5 in 
that window, where $S$ is the number of signal events (which is also 
required to exceed 10) and $B$ is the expected 
number of Drell-Yan background 
events. The Drell-Yan background cross-section is shown in Figure 
\ref{fig:DYmumu}.
Trigger and detector efficiencies are not included, but these are 
expected to be very high for high-mass dimuon events, and the QCD $K$-factor 
for the signal is not included. Dimuon backgrounds from sources 
other than Drell-Yan can be suppressed by requiring no extra hard 
jets or missing energy. In the following, I will again assume a 
spin-averaged BR$(\mu^+\mu^-) = 0.093$ for the signal.
There is also a potential confirming signal from annihilation to 
$e^+e^-$, with an invariant mass peak that is similar but wider and 
smaller due to detector resolution and efficiency effects.

For a 1 fb$^{-1}$ LHC run at $\sqrt{s} = 7$ TeV, the signal cross-section 
in Figure \ref{fig:Dquirkprod} yields 20 expected dimuon events 
for $\mu_D = 
500$ GeV, and as shown in Figure \ref{fig:DYmumu} there is about 1 
background event expected in the corresponding mass window 
$M(\mu^+\mu^-)  = 1000 \pm 100$ GeV. Requiring 10 signal events, the 
discovery reach 
is up to about $\mu_D = 550$ GeV.

For LHC $pp$ collisions at $\sqrt{s} = 14$ TeV, the signal 
cross-section times dimuon branching ratio
for $\mu_D = 800$ GeV is 15 fb, with a background level in the mass 
window $M(\mu^+\mu^-) = 1600 \pm 160$ GeV of 0.8 fb. Therefore, discovery may be 
possible in this case with 1 fb$^{-1}$. The mass reach is essentially 
determined by the number of signal events, since the background levels in 
the high-mass windows are small. In the same way, with 10 fb$^{-1}$, 
I estimate the 
10-event discovery reach to be up to $\mu_D = 1200$ GeV, and for 100 
fb$^{-1}$ up to about $\mu_D$ = 1600 GeV. 

In a more pessimistic scenario, the quirk and antiquirk may usually 
annihilate before they can settle into a low-lying color-singlet quirkonium state. 
The branching ratio to dileptons will be severely reduced in that case
because there are color octet as well as color singlet decay states 
available, and the remaining dimuons will be distributed over larger  
invariant masses. If one supposes that only 10\% of the $D \overline D$ 
pairs that are produced will settle into a low-lying 
color-singlet quirkonium state before annihilation, 
and uses only the dimuon events from this quirkonium peak, 
then the signal cross-section before BR$(\mu^+\mu^-)$ is effectively
ten times smaller than shown in Figure 
\ref{fig:Dquirkprod}.
The limit from comparing to the CDF bound 
on cross-section times branching ratio 
(Figure 3 in \cite{Aaltonen:2008ah}, based on 2.3 fb$^{-1}$) 
results in $\mu_D > 180$ GeV. I 
estimate that the expected reach from a 1 fb$^{-1}$ LHC run at $\sqrt{s} 
= 7$ TeV in this more pessimistic case is roughly $\mu_D = 350$ GeV, for 
which about 17 dimuon signal events and 6 background events would be expected in 
a mass window $M(\mu^+\mu^-) = 700 \pm 70$ GeV. For LHC runs 
at $\sqrt{s} = 14$ TeV with (1, 10, 100) fb$^{-1}$, I similarly estimate 
that the discovery reach for $D \overline D$ that 
annihilate at least 10\% of the time from color-singlet 
$S$-wave quirkonium would 
extend to about $\mu_D = (500, 800, 1100)$ GeV.

\begin{figure}[!tp]
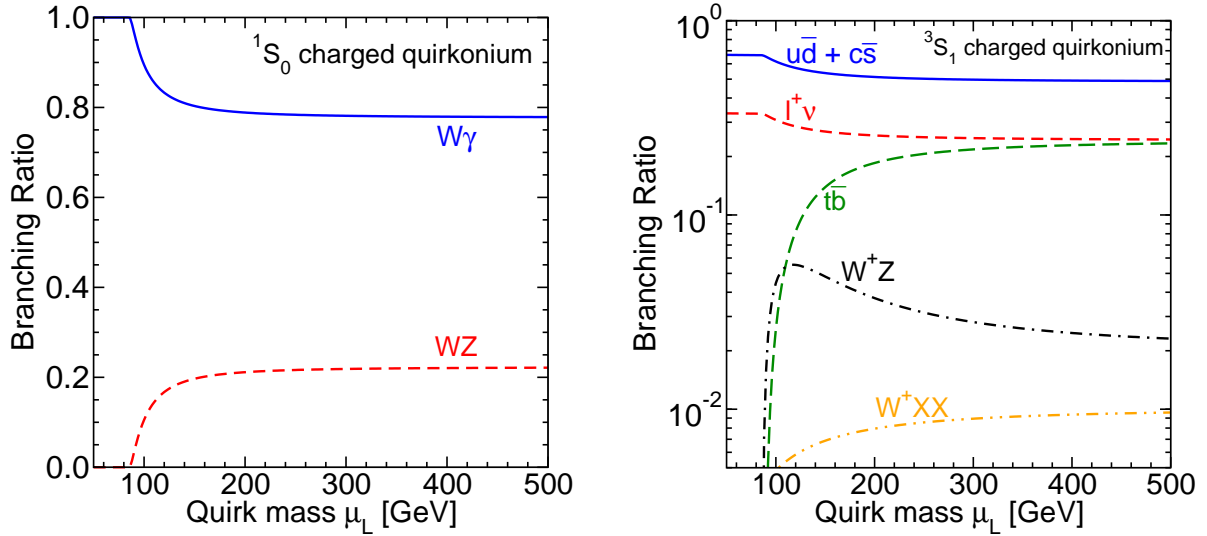

\includegraphics[width=7.5cm,angle=0]{etacBR.eps}
\hspace{0.5cm}
\includegraphics[width=7.5cm,angle=0]{psicBR.eps}
\caption{\label{fig:quirkoniumBRs}
Branching ratios for charged quirkonium  
(a low-lying $\overline L^+ L^0$
or $\overline L^0 L^-$ bound state) in a ${}^1S_0$ (left 
panel) or a
${}^3S_1$ (right panel) state.}
\end{figure}

In the case of the non-colored quirks 
$\overline L^+, L^-, \overline L^0, L^0$, the production rates are 
electroweak, and the energy loss rate for the quirk-antiquirk bound by 
the flux string is much lower \cite{Kang:2008ea}. 
The most promising signal may come from the production of the 
quirk-antiquirk states with a net $\pm 1$ charge, as in 
eqs.~(\ref{eq:proplus}) and (\ref{eq:prominus}), because charge conservation 
then prohibits the subsequent prompt annihilation to invisible $G_X$ 
glueballs that may occur in the case of neutral bound states. The analogous
case for fractionally charged squirks in ``folded supersymmetry" was proposed and 
studied in \cite{Burdman:2008ek}. 
The excess energy from the hard production
will be radiated away in the form of $G_X$ glueballs or soft 
photons, hopefully allowing the quirk and antiquirk to
finally annihilate when nearly at rest in a charged quirkonium bound 
state. The 
annihilation 
is strongest in an $S$-wave state. 
I will again assume that spins are randomized by the energy loss 
process, so that ${}^3S_1$ and 
${}^1S_0$ states are populated in 
the ratio of 3 to 1.
The 
branching ratios for such states have been 
computed in ref.~\cite{Cheung:2008ke}, and are shown in Figure 
\ref{fig:quirkoniumBRs}
for the present case of constituent quirks with charges $\pm 1$ and $0$.
The ${}^1S_0$ state decays predominantly into $W\gamma$, with an 
invariant mass of nearly $2 \mu_L$, and therefore a hard photon. 
(A somewhat smaller branching ratio to $W\gamma$ was obtained in 
ref.~\cite{Cheung:2008ke} for a case with fractionally charged constituent quirks.)
This 
state may therefore be searched for in the $\ell^\pm \gamma + \missET$ 
channel at hadron colliders, as suggested in the similar squirk case 
of ref.~\cite{Burdman:2008ek}.
 
\begin{figure}[!tp]
\begin{minipage}[]{0.52\linewidth}
\begin{flushleft}
\includegraphics[width=7.5cm,angle=0]{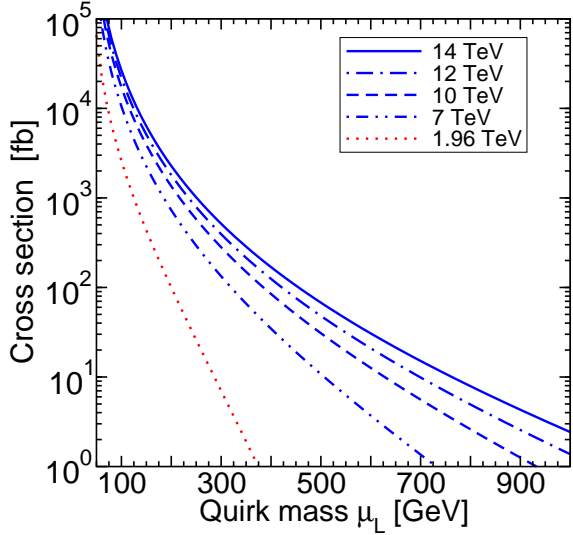}
\end{flushleft}
\end{minipage}
\begin{minipage}[]{0.47\linewidth}
\caption{\label{fig:sigmacharged}
Combined leading-order production cross-section for $\overline L^+ L^0$
and $\overline L^0 L^-$ in $pp$ collisions at the LHC for $\sqrt{s} = 14$, 
12, 10, and 7 TeV, and in $p \overline p$ collisions at the Tevatron 
with $\sqrt{s} = 1.96$ TeV. 
CTEQ5LO \cite{CTEQ5} PDFs were used with $Q=\mu_L$.} \end{minipage}
\end{figure}

The combined\footnote{The 
charge $+1$ combination is produced more often than the 
charge $-1$ one at the LHC, as usual.} production cross-sections at the 
Tevatron and at various possible LHC energies for the charged 
quirk-antiquirk combination are shown in Figure \ref{fig:sigmacharged}. 
These cross-sections are about an order of magnitude larger than for 
fractionally-charged scalar quirks (as studied in 
ref.~\cite{Burdman:2008ek}) of the same mass.
Partly counteracting this, one might expect that
only about 1/4 of fermionic quirk-antiquirk production will end up in a 
${}^1S_0$ state that can annihilate to $W\gamma$, rather than a ${}^3S_1$ 
state that decays mostly to jets or a single lepton plus neutrino. Thus, 
the effective branching ratio of charged quirkonium should be about $0.2$ 
for $W\gamma$, a factor of 3-4 smaller than used in 
\cite{Burdman:2008ek}. The net effect is that the total production 
cross-section times branching ratio for $W\gamma$ should be a factor of 
2-3 times larger, for a given quirkonium mass, than in the study 
of ref.~\cite{Burdman:2008ek}.

The largest background is from Standard Model $W\gamma$ production, which 
features a rapidly falling tail at high photon $p_T$. In contrast, the 
signal from ${}^1S_0$ quirkonium decaying to $W\gamma$ should have a 
photon $p_T$ distribution that is approximately flat, with an endpoint 
near $\mu_L - m_W^2/4 \mu_L$ in the idealized case that the 
transverse kick to the quirkonium is small. 
The relevant photon $p_T$ distribution 
has been studied at Tevatron by both CDF \cite{CDFWgamma} and D$\emptyset$
\cite{DzeroWgamma}, where it was found 
that the data is described well by the SM $W\gamma$ and other subdominant 
backgrounds including $Wj$ with the jet faking a photon and $Z\gamma$ 
with one lepton from the $Z$ missed. At hadron colliders, a $W\gamma$ 
mass peak can in principle be reconstructed if one assumes that the 
observed $\missET$ in the event is due to the neutrino from the leptonic 
$W$ decay, but this is subject to the considerable uncertainty in how 
much missing energy is actually due to missing $G_X$-glueballs  
radiated from the initial state, as well as from 
the underlying event, additional jets, or 
from mismeasurement. However, the discovery potential may be greatly 
enhanced because one can also look for a large number of soft photons 
radiated as the quirk-antiquirk flux string loses energy, forming an 
anomalous ``underlying event" with distinctive character. The resulting 
complications are beyond the scope of the present paper, but have been 
discussed in the analogous case of fractionally charged colorless squirks 
in \cite{Burdman:2008ek,Harnik:2008ax}. The search for $W\gamma$ 
candidates a with large photon $p_T$ and a possible peak in invariant 
mass or transverse mass, in combination with an anomalous underlying 
event used as a background-reducing tag, may well be the best hope 
to detect the quirks in these models.

\section{Conclusions}
\label{sec:outlook}
\setcounter{equation}{0}
\setcounter{footnote}{1}

Extensions of minimal supersymmetry with an extra non-Abelian gauge group 
and quirk supermultiplets maintain two of the hallmark successes of the 
MSSM: compatibility with perturbative gauge coupling unification and with 
constraints on precision electroweak observables. Natural mechanisms can 
put the quirk fermion masses at the TeV scale or below. It follows that 
requiring the unified gauge couplings to be perturbative, so that 
low-scale predictivity is not lost and the apparent unification of gauge 
couplings is not just an accident, the gauge group under which the new 
vector-like particles transform in the fundamental representation must be 
either $SU(2)$, $SU(3)$, or $SO(3)$. The presence of a new non-Abelian 
gauge group has a dramatic effect on the superpartner mass spectrum, and 
allows the soft supersymmetry breaking parameters to be dominated by the 
gaugino masses at the unification scale, while still having a neutralino 
LSP.

In the $SO(3)$ case, the confinement length scale is so large as to make 
the new particles essentially free in collider experiments. In contrast, 
in the minimal versions of the $SU(2)$ and $SU(3)$ cases, the 
quirk-antiquirk bound states produced at colliders will be microscopic. 
If a significant fraction of the colored quirk-antiquirk pairs produced 
at hadron colliders will lose most of their excess energy before 
annihilating as quirkonium, then there is significant reach in the 
dilepton mass peak channel. Even if this fraction is only 10\%, Tevatron 
data that has already been analyzed should allow a limit of $180$ GeV for 
the quirk mass to be set from the search for a $\mu^+\mu^-$ resonance. In 
the optimistic idealized case that all of the quirk-antiquirk pairs 
annihilate from quirkonium, this limit should be about 375 GeV. In a 1 
fb$^{-1}$ LHC run at $\sqrt{s} = 7$ TeV, the discovery reach could be as 
high as 550 GeV, and should extend well above 1 TeV with 100 fb$^{-1}$ at 
$\sqrt{s} = 14$ TeV. The color singlet quirks in these models can also be 
searched for as quirkonium $W\gamma$ resonances. In all cases, the 
quirkonium peak can be accompanied by an anomalous underlying event 
consisting of many soft pions or photons, which can significantly aid in 
making a discovery \cite{Kang:2008ea,Harnik:2008ax}. If low-energy 
supersymmetry is realized in nature, then it will be important to test 
the possibility that it is not minimal by searching for these events.

\bigskip \noindent 
{\it Acknowledgments:} 
I am grateful to Ricky Fok, Roni Harnik, Jos\'e Juknevich, and Graham 
Kribs for helpful communications. This work was supported in part by the 
National Science Foundation grant number PHY-0757325.



\begin{thebibliography}{90}
\baselineskip=12.7pt

\bibitem{Peskin:1991sw}
  M.E.~Peskin and T.~Takeuchi,
  Phys.\ Rev.\ Lett.\  {\bf 65}, 964 (1990),
  Phys.\ Rev.\  D {\bf 46}, 381 (1992).

\bibitem{Golden:1990ig}
  M.~Golden and L.~Randall,
  Nucl.\ Phys.\  B {\bf 361}, 3 (1991).

\bibitem{Holdom:1990tc}
  B.~Holdom and J.~Terning,
  Phys.\ Lett.\  B {\bf 247}, 88 (1990).

 \bibitem{Marciano:1990dp}
  W.J.~Marciano and J.L.~Rosner,
  Phys.\ Rev.\ Lett.\  {\bf 65}, 2963 (1990)
  [Erratum-ibid.\  {\bf 68}, 898 (1992)].

\bibitem{Kennedy:1990ib}
  D.C.~Kennedy and P.~Langacker,
  Phys.\ Rev.\ Lett.\  {\bf 65}, 2967 (1990)
  [Erratum-ibid.\  {\bf 66}, 395 (1991)].
  Phys.\ Rev.\  D {\bf 44}, 1591 (1991).

\bibitem{Altarelli:1990zd}
  G.~Altarelli and R.~Barbieri,
  Phys.\ Lett.\  B {\bf 253}, 161 (1991).
  G.~Altarelli, R.~Barbieri and S.~Jadach,
  Nucl.\ Phys.\  B {\bf 369}, 3 (1992)
  [Erratum-ibid.\  B {\bf 376}, 444 (1992)].

\bibitem{primer}
  For a review, S.P.~Martin,
  ``A supersymmetry primer,''
  [hep-ph/9709356] (version 5, December 2008).

\bibitem{Babu:2004xg}
  K.S.~Babu, I.~Gogoladze and C.~Kolda,
  ``Perturbative unification and Higgs boson mass bounds,''
  [hep-ph/0410085].

\bibitem{Moroi:1991mg}
  T.~Moroi and Y.~Okada,
  Mod.\ Phys.\ Lett.\  A {\bf 7}, 187 (1992),
  Phys.\ Lett.\  B {\bf 295}, 73 (1992).

\bibitem{Babu:2008ge}
  K.S.~Babu, I.~Gogoladze, M.U.~Rehman and Q.~Shafi,
  Phys.\ Rev.\  D {\bf 78}, 055017 (2008)
  [hep-ph/0807.3055].

\bibitem{Martin:2009bg}
  S.P.~Martin,
  Phys.\ Rev.\  D {\bf 81}, 035004 (2010)
  [hep-ph/0910.2732].

\bibitem{Graham:2009gy}  
  P.W.~Graham, A.~Ismail, S.~Rajendran and P.~Saraswat,
  Phys.\ Rev.\  D {\bf 81}, 055016 (2010)
  [hep-ph/0910.3020].  

\bibitem{Martin:2010dc}
  S.P.~Martin,
  Phys.\ Rev.\  {\bf D82}, 055019 (2010).
  [hep-ph/1006.4186].

\bibitem{othervectorlike}
C.~Liu,
  Phys.\ Rev.\  D {\bf 80}, 035004 (2009)
  [hep-ph/0907.3011].
T.~Ibrahim and P.~Nath,
  Phys.\ Rev.\  D {\bf 81}, 033007 (2010)
  [hep-ph/1001.0231].
I.~Gogoladze, Y.~Mimura, N.~Okada and Q.~Shafi,
  Phys.\ Lett.\  B {\bf 686}, 233 (2010)
  [hep-ph/1001.5260].
T.~Li and D.~V.~Nanopoulos,
  [hep-ph/1005.3798].
N.~Craig and J.~March-Russell,
  [hep-ph/1007.0019].
T.~Ibrahim and P.~Nath,
  Phys.\ Rev.\  D {\bf 82}, 055001 (2010)  [hep-ph/1007.0432].
I.~Donkin and A.~Hebecker,
  JHEP {\bf 1009}, 044 (2010)
  [hep-ph/1007.3990].

\bibitem{Okun:1980kw}  
  L.B.~Okun,
  JETP Lett.\  {\bf 31}, 144 (1980)
  [Pisma Zh.\ Eksp.\ Teor.\ Fiz.\  {\bf 31}, 156 (1979)],
  Nucl.\ Phys.\  B {\bf 173}, 1 (1980).

\bibitem{Gupta:1981ve}
  S.~Gupta and H.~R.~Quinn,
  Phys.\ Rev.\  D {\bf 25}, 838 (1982).

\bibitem{Strassler:2006im}
  M.J.~Strassler and K.M.~Zurek,
  Phys.\ Lett.\  B {\bf 651}, 374 (2007)
  [hep-ph/0604261].

\bibitem{Kang:2008ea}
  J.~Kang and M.A.~Luty,
  JHEP {\bf 0911}, 065 (2009)
  [hep-ph/0805.4642].

\bibitem{Burdman:2008ek}
  G.~Burdman, Z.~Chacko, H.S.~Goh, R.~Harnik and C.A.~Krenke,
  Phys.\ Rev.\  D {\bf 78}, 075028 (2008)
  [hep-ph/0805.4667].

\bibitem{Cheung:2008ke}
  K.~Cheung, W.Y.~Keung and T.C.~Yuan,
  Nucl.\ Phys.\  B {\bf 811}, 274 (2009)
  [hep-ph/0810.1524].

\bibitem{Harnik:2008ax}
  R.~Harnik and T.~Wizansky,
  Phys.\ Rev.\  D {\bf 80}, 075015 (2009)
  [hep-ph/0810.3948].

\bibitem{Cai:2008au}
  H.~Cai, H.C.~Cheng and J.~Terning,
  JHEP {\bf 0905}, 045 (2009)
  [hep-ph/0812.0843].

\bibitem{Juknevich:2009ji}
  J.E.~Juknevich, D.~Melnikov and M.J.~Strassler,
  JHEP {\bf 0907}, 055 (2009)
  [hep-ph/0903.0883].

\bibitem{Falkowski:2009yz}
  A.~Falkowski, J.~Juknevich and J.~Shelton,
  ``Dark Matter Through the Neutrino Portal,''
  [hep-ph/0908.1790].

\bibitem{Kribs:2009fy}
  G.D.~Kribs, T.S.~Roy, J.~Terning and K.M.~Zurek,
  Phys.\ Rev.\  D {\bf 81}, 095001 (2010)
  [hep-ph/0909.2034].

\bibitem{Juknevich:2009gg}
  J.E.~Juknevich,
  JHEP {\bf 1008}, 121 (2010)
  [hep-ph/0911.5616].

\bibitem{Abazov:2010yb}
  V.M.~Abazov {\it et al.}  [D$\emptyset$ Collaboration],
  ``Search for new fermions ('quirks') at the Fermilab Tevatron Collider,''
  [hep-ex/1008.3547].

\bibitem{betas:1}
D.R.T.~Jones,
  Nucl.\ Phys.\  B {\bf 87}, 127 (1975).
D.R.T.~Jones and L.~Mezincescu,
  Phys.\ Lett.\  B {\bf 136}, 242 (1984).
P.C.~West,
  Phys.\ Lett.\  B {\bf 137}, 371 (1984).
A.~Parkes and P.C.~West,
  Phys.\ Lett.\  B {\bf 138}, 99 (1984).

\bibitem{betas:2}
S.P.~Martin and M.T.~Vaughn,
  Phys.\ Rev.\  D {\bf 50}, 2282 (1994),
  Erratum Phys.\ Rev.\ D {\bf 78}, 039903 (2008).
  [hep-ph/9311340].
Y.~Yamada,
  Phys.\ Rev.\  D {\bf 50}, 3537 (1994)
  [hep-ph/9401241].
I.~Jack and D.R.T.~Jones,
  Phys.\ Lett.\  B {\bf 333}, 372 (1994)
  [hep-ph/9405233].
I.~Jack et al,
  Phys.\ Rev.\  D {\bf 50}, 5481 (1994)
  [hep-ph/9407291].

\bibitem{Kim:1983dt}
  J.E.~Kim and H.P.~Nilles,
  Phys.\ Lett.\  B {\bf 138}, 150 (1984).

\bibitem{Murayama:1992dj}  
  H.~Murayama, H.~Suzuki and T.~Yanagida,
  Phys.\ Lett.\  B {\bf 291}, 418 (1992).

\bibitem{minX}
S.P.~Martin,
  Phys.\ Rev.\  D {\bf 54}, 2340 (1996) [hep-ph/9602349],
  Phys.\ Rev.\  D {\bf 61}, 035004 (2000)
  [hep-ph/9907550].
  Phys.\ Rev.\  D {\bf 62}, 095008 (2000)
  [hep-ph/0005116].

\bibitem{Chetyrkin:1997un}
  K.G.~Chetyrkin, B.A.~Kniehl and M.~Steinhauser,
  Nucl.\ Phys.\  B {\bf 510}, 61 (1998)  
  [hep-ph/9708255].

\bibitem{fourloop}
  T.~van Ritbergen, J.A.M.~Vermaseren and S.A.~Larin,
  Phys.\ Lett.\  B {\bf 400}, 379 (1997)
  [hep-ph/9701390],
  M.~Czakon,
  Nucl.\ Phys.\  B {\bf 710}, 485 (2005)
  [hep-ph/0411261].

\bibitem{PRH}
  B.~Pendleton and G.G.~Ross,
  Phys.\ Lett.\  B {\bf 98}, 291 (1981).
  C.T.~Hill,
  Phys.\ Rev.\  D {\bf 24}, 691 (1981).

\bibitem{MV}
  S.P.~Martin and M.T.~Vaughn,
  Phys.\ Lett.\  B {\bf 318}, 331 (1993)
  [hep-ph/9308222].

\bibitem{PBMZ}
  D.M.~Pierce, J.A.~Bagger, K.T.~Matchev and R.j.~Zhang,
  Nucl.\ Phys.\  B {\bf 491}, 3 (1997)
  [hep-ph/9606211].

\bibitem{MV2}
Multi-loop contributions to the gluino pole mass are found in 
  Y.~Yamada,
  Phys.\ Lett.\  B {\bf 623}, 104 (2005)
  [hep-ph/0506262],
  S.P.~Martin,
  Phys.\ Rev.\  D {\bf 72}, 096008 (2005)
  [hep-ph/0509115],
  Phys.\ Rev.\  D {\bf 74}, 075009 (2006)
  [hep-ph/0608026].

\bibitem{Kane:1998im}
  G.L.~Kane, S.F.~King,
  Phys.\ Lett.\  {\bf B451}, 113-122 (1999).
  [hep-ph/9810374],
M.~Bastero-Gil, G.L.~Kane, S.F.~King,
  Phys.\ Lett.\  {\bf B474}, 103-112 (2000).
  [hep-ph/9910506].

\bibitem{Chan:1997bi}
  K.L.~Chan, U.~Chattopadhyay, P.~Nath,
  Phys.\ Rev.\  {\bf D58}, 096004 (1998).
  [hep-ph/9710473].

\bibitem{Feng:1999mn}
J.L.~Feng, K.T.~Matchev, T.~Moroi,
  Phys.\ Rev.\ Lett.\  {\bf 84}, 2322-2325 (2000).
  [hep-ph/9908309],
J.L.~Feng, K.T.~Matchev, T.~Moroi,
  Phys.\ Rev.\  {\bf D61}, 075005 (2000).
 [hep-ph/9909334].
J.L.~Feng, K.T.~Matchev, F.~Wilczek,
  Phys.\ Lett.\  {\bf B482}, 388-399 (2000).
  [hep-ph/0004043].

\bibitem{Thomas:1998wy}
  S.D.~Thomas and J.D.~Wells,
  Phys.\ Rev.\ Lett.\  {\bf 81}, 34 (1998)
  [hep-ph/9804359].


\bibitem{Kang:2006yd}
  J.~Kang, M.A.~Luty, S.~Nasri,
  JHEP {\bf 0809}, 086 (2008).
  [hep-ph/0611322].

\bibitem{Jacoby:2007nw}
  C.~Jacoby and S.~Nussinov,
  ``The Relic Abundance of Massive Colored Particles after a Late Hadronic
  Annihilation Stage,''
  [hep-ph/0712.2681],
%
  ``Some Comments on the `Quirks' Scenario,''
  [hep-ph/0907.4932].

\bibitem{Allton:2008ty}
  C.~Allton, M.~Teper and A.~Trivini,
  JHEP {\bf 0807}, 021 (2008)
  [hep-lat/0803.1092].

\bibitem{Abe:1989es}
  F.~Abe {\it et al.}  [CDF Collaboration],
  Phys.\ Rev.\ Lett.\  {\bf 63}, 1447 (1989).
%
  M.~Drees and X.~Tata,
  Phys.\ Lett.\  B {\bf 252}, 695 (1990).
%
  A.~Nisati, S.~Petrarca and G.~Salvini,
  Mod.\ Phys.\ Lett.\  A {\bf 12}, 2213 (1997)
  [hep-ph/9707376].
%
  H.~Baer, K.m.~Cheung and J.F.~Gunion,
  Phys.\ Rev.\  D {\bf 59}, 075002 (1999)
  [hep-ph/9806361].
%
  S.~Raby and K.~Tobe,
  Nucl.\ Phys.\  B {\bf 539}, 3 (1999)
  [hep-ph/9807281].
%
  G.~Polesello and A.~Rimoldi,
  ``Reconstruction of quasi-stable charged sleptons in the ATLAS muon
  spectrometer", ATLAS Note ATL-MUON-99-006 (1999).
%
  R.L.~Culbertson {\it et al.}  [SUSY Working Group Collaboration],
  ``Low scale and gauge mediated supersymmetry breaking at the Fermilab
  Tevatron Run II,''
  [hep-ph/0008070].
%
  S.~Ambrosanio et al,
  ``SUSY long-lived massive particles: Detection and physics at the LHC,''
  [hep-ph/0012192].
%
  B.C.~Allanach, C.M.~Harris, M.A.~Parker, P.~Richardson and B.R.~Webber,
  JHEP {\bf 0108}, 051 (2001)
  [hep-ph/0108097].
%
  A.C.~Kraan,
  Eur.\ Phys.\ J.\  C {\bf 37}, 91 (2004)
  [hep-ex/0404001].
%
  W.~Kilian, T.~Plehn, P.~Richardson and E.~Schmidt,
  Eur.\ Phys.\ J.\  C {\bf 39}, 229 (2005)
  [hep-ph/0408088].
%
  J.L.~Hewett, B.~Lillie, M.~Masip and T.G.~Rizzo,
  JHEP {\bf 0409}, 070 (2004)
  [hep-ph/0408248].
%
  K.~Cheung and W.Y.~Keung,
  Phys.\ Rev.\  D {\bf 71}, 015015 (2005)
  [hep-ph/0408335].
%
  S. Hellman, M. Johansen and D. Milstead,
  ``Mass Measurements of R-hadrons with the ATLAS Experiment at the LHC",
  ATLAS Note ATL-PHYS-PUB-2006-015, (2005).
%
  A.~Arvanitaki, S.~Dimopoulos, A.~Pierce, S.~Rajendran and J.~G.~Wacker,
  Phys.\ Rev.\  D {\bf 76}, 055007 (2007)
  [hep-ph/0506242].
%
  A.C.~Kraan, J.B.~Hansen and P.~Nevski,
  Eur.\ Phys.\ J.\  C {\bf 49}, 623 (2007)
  [hep-ex/0511014].
%
  M.~Fairbairn et al, 
  Phys.\ Rept.\  {\bf 438}, 1 (2007)
  [hep-ph/0611040].
%
  G.~Aad {\it et al.}  [The ATLAS Collaboration],
  ``Expected Performance of the ATLAS Experiment - Detector, Trigger and
  Physics,''
  [hep-ex/0901.0512].


\bibitem{Teper:1998kw}
  M.J.~Teper,
  ``Glueball masses and other physical properties of SU(N) gauge theories  
  in D = 3+1: A review of lattice results for theorists,''
  [hep-th/9812187].

\bibitem{Morningstar:1999rf}
  C.J.~Morningstar and M.J.~Peardon,
  Phys.\ Rev.\  D {\bf 60}, 034509 (1999)
  [hep-lat/9901004].

\bibitem{Lucini:2001ej}
  B.~Lucini and M.~Teper,
  JHEP {\bf 0106}, 050 (2001)
  [hep-lat/0103027].

\bibitem{Chen:2005mg}
  Y.~Chen {\it et al.},
  Phys.\ Rev.\  D {\bf 73}, 014516 (2006)
  [hep-lat/0510074].

\bibitem{stoponium1}
  M.~Drees and M.~M.~Nojiri,
  Phys.\ Rev.\ Lett.\  {\bf 72}, 2324 (1994)
  [hep-ph/9310209],
  Phys.\ Rev.\  D {\bf 49}, 4595 (1994)
  [hep-ph/9312213].

\bibitem{stoponium2}
  S.P.~Martin,
  Phys.\ Rev.\  D {\bf 77}, 075002 (2008)
  [hep-ph/0801.0237],
  S.P.~Martin and J.E.~Younkin,
  Phys.\ Rev.\  D {\bf 80}, 035026 (2009)
  [hep-ph/0901.4318],
  J.E.~Younkin and S.P.~Martin,
  Phys.\ Rev.\  D {\bf 81}, 055006 (2010)
  [hep-ph/0912.4813].

\bibitem{Lemaire:2006kf}
  M.C.~Lemaire, V.A.~Litvine and H.~Newman,
  ``Search for Randall-Sundrum excitations of gravitons decaying into two
  photons for CMS at LHC,''
  CMS Note 2006-051.

\bibitem{Barger:1987xg}
  V.D.~Barger, 
  {\it et al.}, 
  Phys.\ Rev.\  {\bf D35}, 3366 (1987).

\bibitem{Aaltonen:2008ah}
  T.~Aaltonen {\it et al.}  [CDF Collaboration],
  Phys.\ Rev.\ Lett.\  {\bf 102}, 091805 (2009)
  [hep-ex/0811.0053].

\bibitem{Belotelov:2006bb}
  I.~Belotelov {\it et al.},
  ``Study of Drell-Yan dimuon production with the CMS detector,''
  CMS-NOTE-2006-123.

\bibitem{CTEQ5}
  H.L.~Lai {\it et al.}  [CTEQ Collaboration],
  Eur.\ Phys.\ J.\  C {\bf 12}, 375 (2000)
  [hep-ph/9903282].

\bibitem{CDFWgamma}
D.E.~Acosta {\it et al.}  [CDF Collaboration],
  Phys.\ Rev.\ Lett.\  {\bf 94}, 041803 (2005)
  [hep-ex/0410008],
A.~Abulencia {\it et al.}  [CDF Collaboration],
  Phys.\ Rev.\  D {\bf 75}, 112001 (2007)
  [hep-ex/0702029],
CDF collaboration, ``Measurement of Wgamma Production in
$p\overline p$ Collisions at $\sqrt{s}= 1.96$ TeV",
{{\tt http://www-cdf.fnal.gov/physics/ewk/2007/wgzg/}}
(unpublished).
  
\bibitem{DzeroWgamma}
V.M.~Abazov {\it et al.}  [D$\emptyset$ Collaboration],
  Phys.\ Rev.\  D {\bf 71}, 091108 (2005)
  [hep-ex/0503048],
  Phys.\ Rev.\ Lett.\  {\bf 100}, 241805 (2008)
  [hep-ex/0803.0030].

\end{thebibliography}
\end{document}